# Hybrid Spatio-Temporal Artificial Noise Design for Secure MIMOME-OFDM Systems

Ahmed El Shafie, Zhiguo Ding, *Senior Member, IEEE*, and Naofal Al-Dhahir, *Fellow, IEEE*

*Abstract*—This paper investigates artificial noise (AN) injection along the temporal and spatial dimensions of a legitimate system to secure its transmissions from potential eavesdropping. In particular, we consider a multiple-input multiple-output (MIMO) orthogonal frequency-division multiplexing (OFDM) system in the presence of a multiple-antenna passive eavesdropper and characterize both the secrecy rate and average secrecy rate of the legitimate system. We assume that the legitimate transmitter knows the full channel state information (CSI) of the links connecting it with its legitimate receiver but does not know the instantaneous CSI of the passive eavesdropper. Closed-form expressions for the secrecy rate and average secrecy rate are derived for the asymptotic case of a large number of transmit antennas. We compare the degrading effects of spatial and temporal AN on the eavesdropper's rate. We also investigate the power allocation between the data and the AN and the power allocation between the spatial and the temporal AN. An upper bound on the secrecy rate loss due to the presence of the eavesdropper is also derived, and computer simulations are carried out to demonstrate the performance of our proposed AN scheme.

## I. INTRODUCTION

INFORMATION secrecy is crucial to a wireless communication system due to the broadcast nature of radio-frequency transmissions. Traditionally, secrecy is provided by designing upper layer sophisticated protocols and algorithms. More specifically, the problem of securing the information from malicious eavesdropping nodes has been tackled by relying on the conventional encryption mechanism. To improve the system security, physical (PHY) layer security has been recently recognized as a valuable tool to guarantee information-theoretic confidentiality of messages transmitted over the wireless medium.

Manuscript received April 2, 2016; revised June 7, 2016 and July 14, 2016; accepted July 15, 2016. Date of publication August 15, 2016; date of current version May 12, 2017. This work was supported by the Qatar National Research Fund (a member of Qatar Foundation) under NPRP Grant 6-149-2-058. The statements made herein are solely the responsibility of the authors. The review of this paper was coordinated by Dr. M. Elkashlan.

A. El Shafie and N. Al-Dhahir are with the Department of Electrical Engineering, University of Texas at Dallas, Dallas, TX 75080 USA (e-mail: ahmed.salahelshafie@gmail.com; aldhahir@utdallas.edu).

Z. Ding is with the School of Computing and Communications, Lancaster University, Lancaster LA1 4YW, U.K. (e-mail: z.ding@lancaster.ac.uk).

Color versions of one or more of the figures in this paper are available online at http://ieeexplore.ieee.org.

Digital Object Identifier 10.1109/TVT.2016.2600255

Perfect secrecy [1] was introduced as the statistical independence between the information bearing message and the eavesdropper's observations. The seminal works of Wyner [2] and Csiszár and Korner [3] characterized the secrecy capacity for a wiretap channel as the maximal rate at which information can be transmitted over the wireless channel while guaranteeing a vanishing mutual information leakage per channel use. Based on these works, the secrecy capacity has been investigated for different channel models and network settings, and more details can be found in [4]–[11].

In recent years, orthogonal frequency-division multiplexing (OFDM) has been used widely in the PHY layer of most wireless and wired communication standards due to its efficiency in converting frequency-selective fading channels to frequency-flat fading subchannels, resulting in high performance at practical complexity. Hence, it is important to investigate PHY-layer security for OFDM systems. In the information-theoretic security literature, OFDM has usually been modeled as a set of parallel Gaussian channels [12]. In [12], Li *et al.* derived the secrecy capacity and developed the corresponding power-allocation schemes for OFDM systems. In [13], the system security was defined in terms of minimum mean-squared error at the eavesdropper. In [14], the OFDM wiretap channel was treated as a special instance of the MIMO wiretap channel, and its secrecy rates were studied by using both Gaussian inputs and quadrature amplitude modulation signal constellations through asymptotic high- and low-signal-to-noise ratio (SNR) analyses.

Recently, many papers studied precoding for PHY-layer security. For example, linear precoding was studied in [15]–[20]. In [15], with the full channel state information (CSI) knowledge at the legitimate transmitter, the authors proposed a Vandermonde precoding scheme that enables the legitimate transmitter to transmit the information signal in the null space of the equivalent eavesdropper multiple-input multiple-output (MIMO) channel matrix. In [16], Tsai and Poor investigated the optimal power-allocation scheme for artificial noise (AN) secure precoding systems. It was assumed that the legitimate transmitter has perfect CSI of the link to its legitimate receiver and knows only the statistics of the potential eavesdropper's CSI. In [17], Swindlehurst studied the MIMO wiretap problem, in which three multiple-antenna nodes share the channel. The author proposed using enough power to guarantee a certain quality of service for the legitimate destination measured by a predefined signal-to-interference-plus-noise ratio (SINR) for successful decoding, and then allocating the remaining power to generate AN to confuse the eavesdropper. Note that, as explained in [5], using the constraints of the bit error rate, mean-square error, or SINR at eavesdropping nodes does not satisfy either weak or strong secrecy requirements, but it often simplifies system design and analysis.

In [21], Qin *et al.* proposed temporal AN injection for the single-input single-output single-antenna eavesdropper OFDM system, in which a time-domain AN signal is added to the data signal before transmission. The temporal AN signal is designed to be canceled at the legitimate receiver prior to data decoding. It was demonstrated that injecting the AN signal with the data in the frequency domain is not beneficial, since it degrades the secrecy rate. In [22], Akitaya *et al.* proposed a temporal AN injection scheme in the time domain for the multiple-input multiple-output multiantenna eavesdropper (MIMOME) OFDM system. It was assumed that the precoders are used over the available subcarriers, which couple the subcarriers and complicates both encoding and decoding processes at the legitimate nodes.

Both spatial AN-aided (in a non-OFDM system) [16], [17] and temporal AN-aided [21], [22] PHY-layer security schemes have been proposed in the literature. However, we are not aware of any existing work that compares both approaches in terms of average secrecy rates and implementation feasibility and complexity. Our goal in this paper is to answer the following two questions.

1) Under what scenarios is spatial AN preferable over temporal AN and vice versa?
2) Under a total AN average power constraint, is there any advantage in the average secrecy rate for a hybrid spatial–temporal AN scheme over a purely temporal or spatial AN scheme?

To answer these questions, we propose a novel hybrid spatial–temporal AN-aided PHY-layer security scheme for MIMOME channels, which is parameterized by the fraction of the total AN power allocated to each type of AN signals. In addition, we analyze the average secrecy rate achieved by this hybrid scheme and derive tight asymptotic bounds on the average secrecy rate as the number of transmit antennas at the legitimate transmitter becomes large.

In this paper, we consider a general scenario with a hybrid spatio-temporal AN injection scheme. In particular, we consider the MIMOME-OFDM wiretap channel, where each node is equipped with multiple antennas and the legitimate transmitter (Alice) has perfect knowledge of the CSI for the wireless links to her legitimate receiver (Bob) only, while the eavesdropper (Eve) has perfect CSI knowledge of all the links in the network. This global CSI assumption at Eve represents the best-case scenario for Eve (the worst case for Alice/Bob) since she knows all the channels between Alice and Bob as well as the used data and AN precoders at Alice.

The main contributions of this paper are summarized as follows.

1) We propose a novel hybrid spatial and temporal AN design. In our proposed scheme, we assume that the data are encoded in the frequency domain, and then, spatial AN is added to the precoded data and injected into a direction orthogonal to the data vector. Prior to each OFDM block transmission, Alice adds the precoded temporal AN symbols to the signal containing both data and spatial AN to confuse Eve.
2) We exploit the spatial and temporal degrees of freedom provided by the available antennas and by the cyclic prefix (CP) structure of OFDM blocks, respectively, to confuse Eve, which effectively increase the secrecy rate of the legitimate system.
3) We derive several new analytic results and insights. We prove that the average secrecy rate is the same, whether Alice uses spatial-only AN, temporal-only AN, or a hybrid AN scheme when the length of CP is much smaller than the OFDM block size. The asymptotic case with infinite antennas at Alice is also investigated. We derive the optimal power-allocation policy for dynamically assigning power between the information-bearing signals and the AN, where the optimal solution for the average secrecy rate maximization problem is obtained.
4) We show analytically and numerically that Eve can perform per-subcarrier decoding instead of computationally expensive joint decoding without significant reduction in her rate.

*Notation*: Unless otherwise stated, lower- and upper-case bold letters denote vectors and matrices, respectively. Lower- and upper-case letters denote time-domain and frequency-domain signals, respectively. $\mathbf{I}_N$ and $\mathbf{F}$ denote, respectively, the identity matrix whose size is $N \times N$ and the fast Fourier transform (FFT) matrix. $\mathbb{C}^{M \times N}$ denotes the set of all complex matrices of size $M \times N$. $(\cdot)^\top$ and $(\cdot)^*$ denote transpose and Hermitian (i.e., complex-conjugate transpose) operations, respectively. $\mathbb{R}^{M \times M}$ denotes the set of real matrices of size $M \times M$. $\|\cdot\|$ denotes the Euclidean norm of a vector. $[\cdot]_{k,l}$ denotes the $(k,l)$th entry of a matrix, and $[\cdot]_k$ denotes the $k$th entry of a vector. The function $\min(\cdot,\cdot)$ ($\max(\cdot,\cdot)$) returns the minimum (maximum) among the values enclosed in brackets. $\text{blkdiag}(\mathbf{A}_1, \ldots, \mathbf{A}_j, \ldots, \mathbf{A}_M)$ denotes a block diagonal matrix where the enclosed elements are the diagonal blocks. $\mathbb{E}[\cdot]$ denotes statistical expectation. $(\cdot)^{-1}$ is the inverse of the matrix in brackets. $\mathbf{0}$ denotes the all-zero matrix and its size is understood from the context. $\otimes$ is the Kronecker product. $\text{Trace}\{\cdot\}$ denotes the sum of the diagonal entries of enclosed matrix in brackets.

## II. SYSTEM MODEL AND ARTIFICIAL NOISE DESIGN

### A. System Model and Assumptions

The investigated transmission scenario assumes one legitimate transmitter (Alice), one legitimate receiver (Bob), and one passive eavesdropper (Eve). For each OFDM block, Alice transmits over $N$ orthogonal subcarriers by sending $N_s$ streams per subcarrier. We assume that the channel matrices remain constant for the whole transmission duration of $\mathcal{M}$ OFDM blocks, where $\mathcal{M}$ is the coherence time. Alice converts the frequency-domain signals to time-domain signals using an $N$-point inverse FFT (IFFT), and adds a CP of $N_{cp}$ samples to the head of every OFDM block. We assume that the CP length is longer than the delay spreads of all the channels between Alice and Bob to eliminate interblock interference at Bob. Moreover, we assume that the delay spreads of the Alice–Eve channels are shorter than or equal to the CP length.[1] To simplify the analysis in Section III, we assume that all channels have the same delay spread, denoted by $\nu$. All the channel coefficients are assumed to be independent and identically distributed (i.i.d.) zero-mean circularly symmetric complex Gaussian random variables with variance $\sigma^2_{\mathrm{A-B}}$ for the Alice–Bob links and $\sigma^2_{\mathrm{A-E}}$ for the Alice–Eve links, respectively. The channel coefficients between Alice and Bob are assumed to be known at Alice, Bob, and Eve, and the channel coefficients between Alice and Eve are assumed to be known at Eve only. The number of antennas at Node $\ell \in \{\mathrm{A, B, E}\}$ is $N_\ell$. The thermal noise samples at receiving node $\ell$ are modeled as zero-mean complex circularly symmetric Gaussian random variables with variance $\kappa_\ell$ Watts/Hz, $\ell \in \{\mathrm{B, E}\}$. A description of the key variables and their dimensions is given in Table I.

---
[1]This is a best-case assumption for Eve; otherwise, her rate will be degraded due to interblock and intrablock interference.

TABLE I
LIST OF KEY VARIABLES AND THEIR DIMENSIONS

| Symbol | Description | Symbol | Description |
|---|---|---|---|
| $N$ | # subcarriers | $N_{\mathrm{cp}}$ | CP length |
| $N_{\mathrm{s}}$ | # data streams | $N_\ell$ | # antennas at Node $\ell \in \{\mathrm{A, B, E}\}$ |
| $P$ | Average transmit power budget | $\kappa_\ell$ | AWGN variance at Node $\ell$ |
| $\Gamma_\ell = P/\kappa_\ell$ | Ratio between $P$ and noise variance | $\overline{x}$ and $\tilde{x}$ | $1-x$ and $1+x$, respectively, when $x$ is scalar |
| $\theta$ | Data power fraction | $1-\theta$ | AN power fraction |
| $\alpha$ | Spatial AN power fraction | $1-\alpha$ | Temporal AN power fraction |
| $\sigma^2_{\mathrm{A-B}}$ | Variance of Alice–Bob channels | $\sigma^2_{\mathrm{A-E}}$ | Variance of Alice–Eve channels |
| $\mathbf{y}^\ell \in \mathbb{C}^{N_\ell N \times 1}$ | Received signal vector at Node $\ell$ | $\mathbf{P}_{N_\ell} \in \mathbb{C}^{N_\ell N \times N_\ell N}$ | Permutation matrix |
| $\mathbf{R}^{\mathrm{cp}}_{N_\ell} \in \mathbb{C}^{N_\ell N \times N_\ell (N+N_{\mathrm{cp}})}$ | CP removal matrix at Node $\ell$ | $\mathbf{T}^{\mathrm{cp}}_{N_\ell} \in \mathbb{C}^{N_\ell (N+N_{\mathrm{cp}}) \times N_\ell N}$ | CP insertion matrix |
| $\tilde{\mathbf{H}} \in \mathbb{C}^{N_{\mathrm{B}}(N+N_{\mathrm{cp}}) \times N_{\mathrm{A}}(N+N_{\mathrm{cp}})}$ | CIR matrix of Alice–Bob link | $\mathbf{H} \in \mathbb{C}^{N_{\mathrm{B}} N \times N_{\mathrm{A}} N}$ | Frequency-domain matrix of Alice–Bob link |
| $\tilde{\mathbf{G}} \in \mathbb{C}^{N_{\mathrm{E}}(N+N_{\mathrm{cp}}) \times N_{\mathrm{A}}(N+N_{\mathrm{cp}})}$ | CIR matrix of Alice–Eve link | $\mathbf{G} \in \mathbb{C}^{N_{\mathrm{E}} N \times N_{\mathrm{A}} N}$ | Frequency-domain matrix of Alice–Eve link |
| $\mathbf{B} \in \mathbb{C}^{N_{\mathrm{A}} N \times (N_{\mathrm{A}}-N_{\mathrm{s}}) N}$ | Overall spatial AN precoding matrix | $\mathbf{B}_k \in \mathbb{C}^{N_{\mathrm{A}} \times (N_{\mathrm{A}}-N_{\mathrm{s}})}$ | Spatial AN precoding matrix at Subcarrier $k$ |
| $\mathbf{Q} \in \mathbb{C}^{N_{\mathrm{A}}(N+N_{\mathrm{cp}}) \times (N(N_{\mathrm{A}}-N_{\mathrm{s}})+N_{\mathrm{cp}} N_{\mathrm{A}})}$ | Temporal AN precoding matrix | $\mathbf{A} \in \mathbb{C}^{N_{\mathrm{A}} N \times N_{\mathrm{s}} N}$ | Overall data precoding matrix |
| $\mathbf{A}_k \in \mathbb{C}^{N_{\mathrm{A}} \times N_{\mathrm{s}}}$ | Data precoding matrix at Subcarrier $k$ | $\mathbf{C}^*_{\mathrm{B}} \in \mathbb{C}^{N_{\mathrm{s}} N \times N_{\mathrm{B}} N}$ | Overall receive filter matrix at Bob |
| $\mathbf{C}^*_k \in \mathbb{C}^{N_{\mathrm{s}} \times N_{\mathrm{B}}}$ | Receive filter matrix of Bob | $\mathbf{z}^\ell$ | AWGN vector at Node $\ell$ |

### B. Proposed Hybrid Spatio-temporal AN-Aided Scheme

To enhance the security of her transmissions, Alice exploits both the temporal and spatial dimensions provided by the CP and multiple antennas at both Alice and Bob to generate and transmit spatial and temporal AN symbols, in order to degrade the effective channels between Alice and the passive eavesdropper. Our proposed hybrid AN-aided transmission scheme is summarized as follows (see Fig. 1).
1) Alice computes the singular value decomposition (SVD) of the channel matrix at each subcarrier. Then, she uses the $N_{\mathrm{s}}$ columns of the right singular vectors of the subcarrier channel matrix corresponding to the $N_{\mathrm{s}}$ largest nonzero singular values to precode the data. The remaining right singular vectors are used to precode the spatial AN symbols.
2) Alice performs the IFFT on the data-plus-spatial AN vector and adds the CP.
3) Alice then adds the precoded temporal AN signals to the data-and-spatial AN time-domain vector.

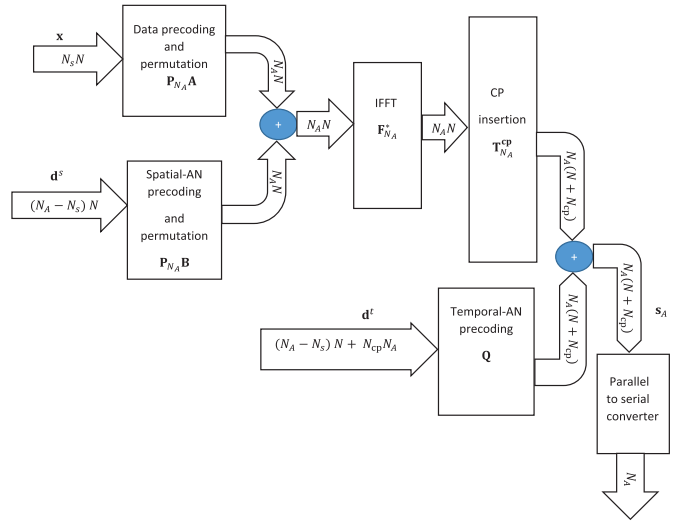

Fig. 1. Block diagram of our proposed hybrid spatial and temporal AN-aided scheme.

As explained in the next subsection, the spatial and temporal AN precoders are designed to ensure that spatial and temporal AN do not harm to Bob.

Define the data symbol permutation matrix $\mathbf{P}_{N_\ell} \in \mathbb{R}^{N_\ell N \times N_\ell N}$, which rearranges the precoded data symbols transmitted/received over the multiple antennas into OFDM blocks. Moreover, define the FFT operation at a node with $N_\ell$ antennas as $\mathbf{F}_{N_\ell} = \mathbf{I}_{N_\ell} \otimes \mathbf{F} \in \mathbb{C}^{N_\ell N \times N_\ell N}$. In addition, Let $\mathbf{d}^s = (\mathbf{d}_1^{s\top}, \mathbf{d}_2^{s\top}, \ldots, \mathbf{d}_N^{s\top})^\top$ be the overall spatial AN vector, and $\mathbf{d}_k^s$ be the injected spatial AN vector over Subcarrier $k$. Moreover, let $\mathbf{d}^t$ be the temporal AN vector. The AN vectors $\mathbf{d}^s$ and $\mathbf{d}^t$ are modeled as complex Gaussian random vectors. As explained in Section II-E, the spatial AN vector is of dimension $N_{\mathrm{A}} - N_{\mathrm{s}}$ and the temporal AN vector is of dimension $N(N_A - N_s) + N_{cp}N_A$. Hence, by neglecting the power of the CP symbols added to the data and spatial-AN signals as in [21] since $N$ is typically much larger than $N_{cp}$, the average transmit power constraint is $\mathbb{E}\{\mathbf{d}^{s*}\mathbf{d}^s\} + \mathbb{E}\{\mathbf{d}^{t*}\mathbf{d}^t\} + \mathbb{E}\{\mathbf{x}^*\mathbf{x}\} \leq P$, where $\mathbf{x} = (\mathbf{x}_1^\top, \mathbf{x}_2^\top, \ldots, \mathbf{x}_N^\top)^\top \in \mathbb{C}^{N_s N \times 1}$ is the data vector, $\mathbf{x}_k \in \mathbb{C}^{N_s \times 1}$ is the data vector transmitted over Subcarrier $k$, and $P$ is the average transmit power budget at Alice in Watts/Hz.

Define $\theta P$ to be the data transmission power; then, we have $\mathbb{E}\{\mathbf{x}^*\mathbf{x}\} = \theta P$ and $\mathbb{E}\{\mathbf{d}^{s*}\mathbf{d}^s\} + \mathbb{E}\{\mathbf{d}^{t*}\mathbf{d}^t\} = \overline{\theta}P$, where $\overline{\theta} = 1 - \theta$. Letting $\alpha$ be the fraction of $\overline{\theta}P$ allocated to spatial AN, then the power allocated to temporal AN is $\overline{\alpha}\overline{\theta}P$, where $\overline{\alpha} = 1 - \alpha$. We assume that the fraction of the power allocated for data and spatial-AN transmissions over Subcarrier $k$ is $1/N$.[2] Hence, the power is allocated equally among the subcarriers.

Since Eve's CSI is unknown at Alice, Alice distributes the power of the spatial and temporal AN uniformly over their precoders' columns. Hence, the power of a spatial AN symbol is $\alpha\overline{\theta}P/((N + N_{cp})(N_A - N_s)) \approx \alpha\overline{\theta}P/(N(N_A - N_s))$, while the power of a temporal AN symbol is $\overline{\alpha}\overline{\theta}P/(N(N_A - N_s) + N_{cp}N_A)$. Therefore, Alice transmits the following vector:

$$\mathbf{s}_A = \mathbf{T}_{N_A}^{cp} \mathbf{F}_{N_A}^* \mathbf{P}_{N_A} (\mathbf{A}\mathbf{x} + \mathbf{B}\mathbf{d}^s) + \mathbf{Q}\mathbf{d}^t \quad (1)$$

where $\mathbf{P}_{N_A} \in \mathbb{R}^{N_A N \times N_A N}$ is a permutation matrix that rearranges the precoded subcarriers,[3] $\mathbf{T}_{N_A}^{cp} \in \mathbb{C}^{N_A(N+N_{cp}) \times N_A N}$ is the CP insertion matrix, $\mathbf{Q} \in \mathbb{C}^{N_A(N+N_{cp}) \times (N(N_A - N_s) + N_{cp}N_A)}$ is the temporal AN precoder matrix, and $\mathbf{A} \in \mathbb{C}^{N_A N \times N_s N}$ and $\mathbf{B} \in \mathbb{C}^{N_A N \times (N_A - N_s)N}$ are the data and spatial AN precoders, respectively.

### C. Received Signal Vector at Bob

After applying the linear filter $\mathbf{C}_B^* \in \mathbb{C}^{N_s N \times N_B N}$, the filtered output vector at Bob is given by

$$\mathbf{C}_B^* \mathbf{y}^B = \mathbf{C}_B^* \mathbf{P}_{N_B}^\top \mathbf{F}_{N_B} \mathbf{R}_{N_B}^{cp} \tilde{\mathbf{H}} \mathbf{s}_A + \mathbf{C}_B^* \mathbf{z}^B \quad (2)$$

where $\mathbf{y}^B$ is the receive signal vector at Bob, and $\mathbf{P}_{N_B}^\top$ is a permutation matrix used by Bob to rearrange the received subcarriers. In addition, $\mathbf{F}_{N_B} = \text{blkdiag}(\mathbf{F}, \mathbf{F}, \ldots, \mathbf{F}) \in \mathbb{C}^{N_B N \times N_B N}$ is the FFT matrix, and $\mathbf{R}_{N_B}^{cp} \in \mathbb{C}^{N_B N \times N_B(N+N_{cp})}$ is the CP removal matrix used at Bob. Finally, $\tilde{\mathbf{H}} \in \mathbb{C}^{N_B(N+N_{cp}) \times N_A(N+N_{cp})}$ is the channel impulse response (CIR) matrix of the Alice–Bob link, and $\mathbf{z}^B \in \mathbb{C}^{N_B N \times 1}$ is the additive white Gaussian noise (AWGN) vector at Bob.

The matrix $\mathbf{H} = \mathbf{P}_{N_B}^\top \mathbf{F}_{N_B} \mathbf{R}_{N_B}^{cp} \tilde{\mathbf{H}} \mathbf{T}_{N_A}^{cp} \mathbf{F}_{N_A}^* \mathbf{P}_{N_A}$, whose size is $N_B N \times N_A N$, is block-diagonal, i.e., $\mathbf{H} = \text{blkdiag}(\mathbf{H}_1, \mathbf{H}_2, \ldots, \mathbf{H}_N)$, where $\mathbf{H}_k \in \mathbb{C}^{N_B \times N_A}$ is the frequency-domain channel matrix of the Alice–Bob link at Subcarrier $k$. The data precoding matrix of the $k$th subcarrier at Alice is denoted by $\mathbf{A}_k \in \mathbb{C}^{N_A \times N_s}$, where $[\mathbf{A}_k]_{i,j}$, for all $j$, represents the weights multiplied by the data symbol $X_{i,k}$. Hence, the overall data precoding matrix is denoted by $\mathbf{A} = \text{blkdiag}(\mathbf{A}_1, \mathbf{A}_2, \ldots, \mathbf{A}_N) \in \mathbb{C}^{N_A N \times N_s N}$.

The spatial AN precoder matrix of Subcarrier $k$ at Alice is $\mathbf{B}_k \in \mathbb{C}^{N_A \times (N_A - N_s)}$, where $[\mathbf{B}_k]_{i,j}$, for all $j$, represents the weights multiplied by the spatial AN symbol $[\mathbf{d}_k^s]_{j,1}$. The overall spatial AN precoding matrix is denoted by $\mathbf{B} = \text{blkdiag}(\mathbf{B}_1, \mathbf{B}_2, \ldots, \mathbf{B}_N) \in \mathbb{C}^{N_A N \times (N_A - N_s)N}$. Moreover, the receive filter matrix applied at Bob over Subcarrier $k$ is denoted by $\mathbf{C}_k^* \in \mathbb{C}^{N_s \times N_B}$. The overall filtering matrix at Bob, denoted by $\mathbf{C}_B^* \in \mathbb{C}^{N_s N \times N_B N}$, is given by $\mathbf{C}_B^* = \text{blkdiag}(\mathbf{C}_1^*, \mathbf{C}_2^*, \ldots, \mathbf{C}_N^*)$.

### D. Design of Alice's Data Precoder Matrix and Bob's Receive Filter Matrix

The received signal at Bob after canceling the temporal and spatial AN vectors is given by

$$\mathbf{C}_B^* \mathbf{y}^B = \begin{pmatrix} \mathbf{C}_1^* \mathbf{y}_1^B \\ \mathbf{C}_2^* \mathbf{y}_2^B \\ \vdots \\ \mathbf{C}_N^* \mathbf{y}_N^B \end{pmatrix} = \mathbf{C}_B^* \mathbf{H} \mathbf{A} \mathbf{x} + \mathbf{C}_B^* \mathbf{z}^B$$

$$= \mathbf{C}_B^* \mathbf{P}_{N_B}^\top \mathbf{F}_{N_B} \mathbf{R}_{N_B}^{cp} \tilde{\mathbf{H}} \mathbf{T}_{N_A}^{cp} \mathbf{F}_{N_A}^* \mathbf{P}_{N_A} \mathbf{A} \mathbf{x} + \mathbf{C}_B^* \mathbf{z}^B \quad (3)$$

where $\mathbf{y}_k^B \in \mathbb{C}^{N_B \times 1}$ is the received vector at Subcarrier $k$ and $\mathbf{C}_k^* \mathbf{y}_k^B = \mathbf{C}_k^* \mathbf{H}_k \mathbf{A}_k \mathbf{x}_k + \mathbf{C}_k^* \mathbf{z}_k^B$.

To maximize the SINR per symbol at Bob, both Alice and Bob use the SVD of the per-subcarrier channel matrix, $\mathbf{H}_k = \mathbf{U}_k \mathbf{\Sigma}_k \mathbf{V}_k^*$, where $\mathbf{\Sigma}_k$ contains the singular values, the columns of $\mathbf{U}_k$ are the left-singular vectors of $\mathbf{H}_k$, and the columns of $\mathbf{V}_k$ are the right-singular vectors of $\mathbf{H}_k$. Hence, Alice selects the data precoding matrix $\mathbf{A}_k$ to be the $N_s$ columns of $\mathbf{V}_k$ and $\mathbf{C}_k$ to be the $N_s$ columns of $\mathbf{U}_k$, which correspond to the $N_s$ largest nonzero singular values of $\mathbf{H}_k$.

### E. Design of Alice's Temporal and Spatial AN Precoders

Our aim is to cancel the spatial and temporal AN at Bob. From (2), the condition to cancel the spatial-AN vector at Bob is given by

$$\mathbf{C}_B^* \mathbf{H} \mathbf{B} = \mathbf{0} \Leftrightarrow \mathbf{C}_k^* \mathbf{H}_k \mathbf{B}_k = \mathbf{0} \quad \forall k \in \{1, \ldots, N\}. \quad (4)$$

Given $\mathbf{C}_k^*$, Alice designs the spatial AN precoding matrix $\mathbf{B}_k$ to lie in the null space of $\mathbf{C}_k^* \mathbf{H}_k \in \mathbb{C}^{N_s \times N_A}$ which is nontrivial only if $N_A > N_s$. If this condition is not satisfied (i.e., $N_A = N_s$), Alice will not be able to inject any spatial AN. The spatial AN precoding matrix is designed to make the AN lie in the subspace spanned by the remaining vectors of $\mathbf{V}_k$. That is, Alice combines the remaining $N_A - N_s$ columns of $\mathbf{V}_k$ using Gaussian random variables each with zero mean and variance $\alpha\overline{\theta}P/(N(N_A - N_s))$.

After designing the data and spatial AN linear precoders and receive filters, Alice designs the temporal AN precoder matrix

---

[2]The power allocated to a subcarrier for data and spatial-AN transmissions should be divided by $N + N_{cp}$, instead of $N$. However, in this work, unless otherwise stated explicitly, we assume that $N$ is much larger than $N_{cp}$ and, hence, $N + N_{cp} \approx N$ and the power spent in CP transmission is negligible.

[3]This matrix is used because Alice rearranges the data as a block of $N_s$ precoded symbols per subcarrier. The multiplication of the permutation matrix and the precoded data rearranges the data into OFDM blocks, where each OFDM block consists of $N$ subcarriers. For instance, OFDM block $j$ spans entries $N(j-1) + 1$ to $Nj$.

$\mathbf{Q}$, based on the knowledge of the receive filter at Bob, according to the following condition:

$$\mathbf{C}_\mathrm{B}^* \mathbf{P}_{N_\mathrm{B}}^\top \mathbf{F}_{N_\mathrm{B}} \mathbf{R}_{N_\mathrm{B}}^\mathrm{cp} \tilde{\mathbf{H}} \mathbf{Q} = \mathbf{0} \quad (5)$$

where $\mathbf{Q}$ lies in the null space of $\mathbf{C}_\mathrm{B}^* \mathbf{P}_{N_\mathrm{B}}^\top \mathbf{F}_{N_\mathrm{B}} \mathbf{R}_{N_\mathrm{B}}^\mathrm{cp} \tilde{\mathbf{H}}$, which has a dimension of $N_\mathrm{s} N \times N_\mathrm{A}(N + N_\mathrm{cp})$. The condition to cancel the temporal AN vector at Bob is given by

$$N_\mathrm{A}(N + N_\mathrm{cp}) > N_\mathrm{s} N \Rightarrow (1 + \frac{N_\mathrm{cp}}{N}) > \frac{N_\mathrm{s}}{N_\mathrm{A}}. \quad (6)$$

This condition is always satisfied since $N_\mathrm{A} \geq N_\mathrm{s}$. In Appendix A, we show how Alice can compute $\mathbf{Q}$ efficiently using the properties of the Toeplitz matrix $\mathbf{R}_{N_\mathrm{B}}^\mathrm{cp} \tilde{\mathbf{H}}$.

### F. Received Signal Vector at Eve

Since Eve has $N_\mathrm{E}$ receive antennas, her received signal vector is given by

$$\mathbf{y}^\mathrm{E} = \mathbf{P}_{N_\mathrm{E}}^\top \mathbf{F}_{N_\mathrm{E}} \mathbf{R}_{N_\mathrm{E}}^\mathrm{cp} \tilde{\mathbf{G}} \mathbf{s}_\mathrm{A} + \mathbf{z}^\mathrm{E} \quad (7)$$

where $\mathbf{y}^\mathrm{E} \in \mathbb{C}^{N_\mathrm{E} N \times 1}$ is the received signal vector at Eve, $\mathbf{F}_{N_\mathrm{E}} \in \mathbb{C}^{N_\mathrm{E} N \times N_\mathrm{E} N}$ is the FFT operation performed at Eve, $\mathbf{R}_{N_\mathrm{E}}^\mathrm{cp} \in \mathbb{C}^{N_\mathrm{E} N \times N_\mathrm{E}(N+N_\mathrm{cp})}$ is the CP removal matrix used at Eve, $\tilde{\mathbf{G}} \in \mathbb{C}^{N_\mathrm{E}(N+N_\mathrm{cp}) \times N_\mathrm{A}(N+N_\mathrm{cp})}$ is the CIR matrix of the Alice–Eve link, and $\mathbf{z}^\mathrm{E} \in \mathbb{C}^{N_\mathrm{E} N \times 1}$ is the AWGN vector at Eve's receiver.

The per-subcarrier received signal vector at Eve, denoted by $\mathbf{y}_k^\mathrm{E} \in \mathbb{C}^{N_\mathrm{E} \times 1}$, is given by

$$\mathbf{y}_k^\mathrm{E} = \mathbf{G}_k (\mathbf{A}_k \mathbf{x}_k + \mathbf{B}_k \mathbf{d}_k^s) + \mathbf{E}_k \mathbf{d}^t + \mathbf{z}_k^\mathrm{E} \quad (8)$$

where $\mathbf{G}_k$ is the frequency-domain channel matrix at Subcarrier $k$ of the Alice–Eve link, $\mathbf{z}_k^\mathrm{E}$ are rows $(k-1)N_\mathrm{E}+1$ to $kN_\mathrm{E}$ of the AWGN vector $\mathbf{z}^\mathrm{E}$, and $\mathbf{E}_k$ consists of the $N_\mathrm{E}$ rows from $(k-1)N_\mathrm{E}+1$ to $kN_\mathrm{E}$ of matrix $\mathbf{E} = \mathbf{P}_{N_\mathrm{E}}^\top \mathbf{F}_{N_\mathrm{E}} \mathbf{R}_{N_\mathrm{E}}^\mathrm{cp} \tilde{\mathbf{G}} \mathbf{Q}$. We define the frequency-domain block-diagonal matrix of the Alice–Eve link as $\mathbf{G} = \mathrm{blkdiag}(\mathbf{G}_1, \mathbf{G}_2, \ldots, \mathbf{G}_N)$.

## III. AVERAGE SECRECY RATE

Our goal in this section is to derive closed-form expressions for the average secrecy rate. Assume that the received signal vector at Node $\ell_2$ due to a transmission from Node $\ell_1$, where $\ell_1, \ell_2 \in \{\mathrm{A, B, E}\}$, is $\mathbf{r} + \mathbf{j} + \mathbf{z}^{\ell_2}$, where $\mathbf{r}$ is the received weighted data vector, $\mathbf{z}^{\ell_2}$ is the AWGN vector at Node $\ell_2$, and $\mathbf{j}$ is a Gaussian interference vector. The instantaneous rate of the $\ell_1 - \ell_2$ link is [16], [23]

$$R_{\ell_1 - \ell_2} = \log_2 \det \left( \mathbf{\Sigma}_\mathbf{r} \left[ \mathbb{E}\left\{ (\mathbf{j} + \mathbf{z}^{\ell_2})(\mathbf{j} + \mathbf{z}^{\ell_2})^* \right\} \right]^{-1} + \mathbf{I}_{N_{\ell_2}} \right) \quad (9)$$

where $\mathbf{\Sigma}_\mathbf{r} = \mathbb{E}\{\mathbf{rr}^*\}$ is the data weighted covariance matrix and $\mathbb{E}\left\{(\mathbf{j}+\mathbf{z}^{\ell_2})(\mathbf{j}+\mathbf{z}^{\ell_2})^*\right\}$ is the noise-plus-interference covariance matrix.

Define the input SNRs of the Alice–Bob and Alice–Eve links as $\Gamma_\mathrm{B} = P/\kappa_\mathrm{B}$ and $\Gamma_\mathrm{E} = P/\kappa_\mathrm{E}$, respectively. Assuming that the transmit power is allocated equally among the $N_\mathrm{s}$ independent data symbols, and using (3) and (9), the rate at Bob, denoted by $R_{\mathrm{A-B}}$, is given by

$$R_{\mathrm{A-B}} = \log_2 \det \left( \theta \Gamma_\mathrm{B} \mathbf{C}_\mathrm{B}^* \mathbf{H} \mathbf{A} \mathbf{\Sigma}_\mathbf{x} (\mathbf{C}_\mathrm{B}^* \mathbf{H} \mathbf{A})^* + \mathbf{I}_{N_\mathrm{s} N} \right)$$
$$= \sum_{k=1}^N \log_2 \det \left( \theta \Gamma_\mathrm{B} \mathbf{C}_k^* \mathbf{H}_k \mathbf{A}_k \mathbf{\Sigma}_{\mathbf{x}_k} (\mathbf{C}_k^* \mathbf{H}_k \mathbf{A}_k)^* + \mathbf{I}_{N_\mathrm{s}} \right) \quad (10)$$

where $\theta P \mathbf{\Sigma}_\mathbf{x}$ is the data covariance matrix with $\mathbf{\Sigma}_\mathbf{x} = \frac{1}{N_\mathrm{s}(N+N_\mathrm{cp})} \mathbf{I}_{N_\mathrm{s} N} \approx \frac{1}{N_\mathrm{s} N} \mathbf{I}_{N_\mathrm{s} N}$ and $\mathbf{\Sigma}_{\mathbf{x}_k} \approx \frac{1}{N_\mathrm{s} N} \mathbf{I}_{N_\mathrm{s}}$.

We consider and compare two decoding approaches at Eve. In the first approach, she decodes the signals on all the subcarriers jointly by taking into account the correlation among the subcarriers due to the temporal AN. In the second approach, Eve decodes the signals at each subcarrier separately assuming that each subcarrier is a MIMO channel. If Eve performs joint processing across all of her subcarriers, using (7), her rate, denoted by $R_{\mathrm{A-E}}$, is given by[4]

$$R_{\mathrm{A-E}} = \log_2 \det \left( \theta \Gamma_\mathrm{E} \mathbf{G} \mathbf{A} \mathbf{\Sigma}_\mathbf{x} (\mathbf{G}\mathbf{A})^* (\mathbf{\Sigma}_{\mathrm{AN}} + \mathbf{I}_{N_\mathrm{E} N})^{-1} + \mathbf{I}_{N_\mathrm{E} N} \right) \quad (11)$$

where

$$\mathbf{\Sigma}_{\mathrm{AN}} = \overline{\theta} \Gamma_\mathrm{E} \left( \frac{\alpha \mathbf{G} \mathbf{B} \mathbf{B}^* \mathbf{G}^*}{N(N_\mathrm{A} - N_\mathrm{s})} + \frac{\overline{\alpha} \mathbf{E} \mathbf{E}^*}{N(N_\mathrm{A} - N_\mathrm{s}) + N_\mathrm{cp} N_\mathrm{A}} \right). \quad (12)$$

The eavesdropper's rate expression in (11) assumes that Eve knows the channels between Alice and Bob, and hence, she knows the used data and AN precoders by Alice. This assumption represents a best-case scenario for Eve.

*Proposition 1:* When $1 - \frac{N_\mathrm{s}}{N_\mathrm{A}} \gg \frac{N_\mathrm{cp}}{N}$, Eve's rate and the secrecy rate are independent of $\alpha$.

*Proof:* See Appendix B. ∎

If Eve performs per-subcarrier processing due to its reduced complexity, her rate is given by

$$R_{\mathrm{A-E}} = \sum_{k=1}^N \left[ \log_2 \det \left( \theta \Gamma_\mathrm{E} \mathbf{G}_k \mathbf{A}_k \mathbf{\Sigma}_{\mathbf{x}_k} (\mathbf{G}_k \mathbf{A}_k)^* \right.\right.$$
$$\left.\left. \times (\mathbf{\Sigma}_{\mathrm{AN},k} + \mathbf{I}_{N_\mathrm{E}})^{-1} + \mathbf{I}_{N_\mathrm{E}} \right) \right] \quad (13)$$

where

$$\mathbf{\Sigma}_{\mathrm{AN},k} = \overline{\theta} \Gamma_\mathrm{E} \left( \frac{\alpha \mathbf{G}_k \mathbf{B}_k \mathbf{B}_k^* \mathbf{G}_k^*}{N(N_\mathrm{A} - N_\mathrm{s})} + \frac{\overline{\alpha} \mathbf{E}_k \mathbf{E}_k^*}{N(N_\mathrm{A} - N_\mathrm{s}) + N_\mathrm{cp} N_\mathrm{A}} \right). \quad (14)$$

The secrecy rate of the legitimate system is given by [5], [16], [25], [26]

$$R_{\mathrm{sec}} = (R_{\mathrm{A-B}} - R_{\mathrm{A-E}})^+ \quad (15)$$

---

[4]This expression represents the best rate for Eve which will be reduced if Eve applies a linear filter such as MMSE or zero-forcing filters as in [24].

where $(\cdot)^+ = \max(\cdot, 0)$. The average secrecy rate of the legitimate system is approximately given by [16], [25], [26]

$$\mathbb{E}\{R_{\text{sec}}\} = \mathbb{E}\{R_{\text{A-B}}\} - \mathbb{E}\{R_{\text{A-E}}\}. \quad (16)$$

The average secrecy rate loss due to eavesdropping, which represents the reduction in the average secrecy rate due to the presence of an eavesdropper (i.e., the difference between the secrecy rate of the Alice–Bob link and the achievable rate of the Alice–Bob link without eavesdropping), denoted by $\mathcal{S}_{\text{Loss}}$, is given by

$$\mathcal{S}_{\text{Loss}} = R_{\text{A-B,noEve}} - R_{\text{sec}} \quad (17)$$

where $R_{\text{A-B,noEve}}$ is the achievable rate at Bob when there is no Eve.

The following relation is useful in our secrecy rate derivations:

$$\mathbf{G}_k \mathbf{G}_k^* = \mathbf{G}_k \mathbf{A}_k (\mathbf{G}_k \mathbf{A}_k)^* + \mathbf{G}_k \mathbf{B}_k (\mathbf{G}_k \mathbf{B}_k)^* \quad (18)$$

where the columns of $\mathbf{A}_k$ are the $N_{\text{s}}$ right singular vectors corresponding to the largest nonzero singular values of $\mathbf{H}_k$ and the columns of $\mathbf{B}_k$ are the remaining $N_{\text{A}} - N_{\text{s}}$ right singular vectors of $\mathbf{H}_k$. Equation (18) can be easily obtained by first noting that $\mathbf{A}_k \mathbf{A}_k^* + \mathbf{B}_k \mathbf{B}_k^* = \mathbf{I}_{N_{\text{A}}}$, since the columns of $\mathbf{A}_k$ and $\mathbf{B}_k$ form orthonormal bases. Then, (18) is obtained by left and right multiplication by $\mathbf{G}_k$ and $\mathbf{G}_k^*$, respectively.

In the following subsection, we investigate the asymptotic average secrecy rate of MIMOME-OFDM systems as $N_{\text{A}} \to \infty$. We derive a lower-bound on the average secrecy rate as well as the average secrecy rate loss due to eavesdropping. Moreover, we investigate the asymptotic scenarios of low and high SNRs at sufficiently large $N_{\text{A}}$.

### A. Asymptotic Average Rates in MIMOME-OFDM Channels

In this subsection, we investigate the asymptotic case of $N_{\text{A}} \to \infty$. Wireless base stations with large numbers of transmit antennas, which are also known as massive MIMO systems, have attracted considerable research attention recently (see, e.g., [27] and [28] and the references therein). It is important to point out that massive MIMO is a key-enabling technology for emerging 5G communication networks.

For MIMOME-OFDM systems, Eve's rate when $N_{\text{A}}$ is very large is given by

$$R_{\text{A-E}} \approx \log_2 \det \left( \theta \Gamma_{\text{E}} \mathbf{G} \mathbf{A} \mathbf{\Sigma_x} (\mathbf{G} \mathbf{A})^* (\mathbf{\Sigma}_{\text{AN}} + \mathbf{I}_{N_{\text{E}} N})^{-1} \right.$$
$$\left. + \mathbf{I}_{N_{\text{E}} N} \right) \quad (19)$$

where

$$\mathbf{\Sigma}_{\text{AN}} \approx$$
$$\overline{\theta} \Gamma_{\text{E}} \left( \frac{\alpha}{N(N_{\text{A}} - N_{\text{s}})} \mathbf{G}\mathbf{G}^* + \frac{\overline{\alpha}}{N(N_{\text{A}} - N_{\text{s}}) + N_{\text{cp}} N_{\text{A}}} \mathbf{E}\mathbf{E}^* \right). \quad (20)$$

Note that we used the fact that $\mathbf{B}\mathbf{B}^* \approx \mathbf{I}_{N(N_{\text{A}} - N_{\text{s}})}$ and $\mathbf{Q}\mathbf{Q}^* \approx \mathbf{I}_{N_{\text{A}}(N+N_{\text{cp}})}$ as $N_{\text{A}} \to \infty$. To analyze the average secrecy rate, we need the statistics of $\mathbf{P}_{N_{\text{E}}}^\top \mathbf{F}_{N_{\text{E}}} \mathbf{R}_{N_{\text{E}}}^{\text{cp}} \tilde{\mathbf{G}}$, $\mathbf{G}_k \mathbf{A}_k \mathbf{A}_k^* \mathbf{G}_k^*$, and $\mathbf{H}_k \mathbf{A}_k \mathbf{A}_k^* \mathbf{H}_k^*$, which are derived in Appendixes C and D.

*Proposition 2:* As $N_{\text{A}} \to \infty$ and when $1 - \frac{N_{\text{s}}}{N_{\text{A}}} \gg \frac{N_{\text{cp}}}{N}$, Eve's rate is independent of $\alpha$ and is given by

$$R_{\text{A-E}} \approx \log_2 \det \left( \theta \Gamma_{\text{E}} \mathbf{G} \mathbf{A} \mathbf{\Sigma_x} (\mathbf{G} \mathbf{A})^* \right.$$
$$\left. \times \left( \frac{\overline{\theta} \Gamma_{\text{E}}}{N(N_{\text{A}} - N_{\text{s}})} \mathbf{G}\mathbf{G}^* + \mathbf{I}_{N_{\text{E}} N} \right)^{-1} + \mathbf{I}_{N_{\text{E}} N} \right). \quad (21)$$

In addition, the secrecy rate is given by

$$R_{\text{sec}} \approx \sum_{k=1}^{N} \log_2 \det \left( \theta \Gamma_{\text{B}} \mathbf{C}_k^* \mathbf{H}_k \mathbf{A}_k \mathbf{\Sigma}_{\mathbf{x}_k} (\mathbf{C}_k^* \mathbf{H}_k \mathbf{A}_k)^* + \mathbf{I}_{N_{\text{s}}} \right)$$
$$- \log_2 \det \left( \theta \Gamma_{\text{E}} \mathbf{G} \mathbf{A} \mathbf{\Sigma_x} (\mathbf{G} \mathbf{A})^* \right.$$
$$\left. \times \left( \frac{\overline{\theta} \Gamma_{\text{E}}}{N(N_{\text{A}} - N_{\text{s}})} \mathbf{G}\mathbf{G}^* + \mathbf{I}_{N_{\text{E}} N} \right)^{-1} + \mathbf{I}_{N_{\text{E}} N} \right). \quad (22)$$

*Proof:* See Appendix E. ∎

*Proposition 3:* As $N_{\text{A}} \to \infty$ and when $1 - \frac{N_{\text{s}}}{N_{\text{A}}} \gg \frac{N_{\text{cp}}}{N}$, per-subcarrier processing at Eve achieves the same rate as that achieved with joint subcarrier processing. Hence, the average secrecy rate is given by

$$R_{\text{sec}} \approx \sum_{k=1}^{N} \left( \log_2 \det \left( \theta \Gamma_{\text{B}} \mathbf{C}_k^* \mathbf{H}_k \mathbf{A}_k \mathbf{\Sigma}_{\mathbf{x}_k} (\mathbf{C}_k^* \mathbf{H}_k \mathbf{A}_k)^* + \mathbf{I}_{N_{\text{s}}} \right) \right.$$
$$- \log_2 \det \left( \theta \Gamma_{\text{E}} \mathbf{G}_k \mathbf{A}_k \mathbf{\Sigma}_{\mathbf{x}_k} (\mathbf{G}_k \mathbf{A}_k)^* \right.$$
$$\left. \left. \times \left( \frac{\overline{\theta} \Gamma_{\text{E}}}{N(N_{\text{A}} - N_{\text{s}})} \mathbf{G}_k \mathbf{G}_k^* + \mathbf{I}_{N_{\text{E}}} \right)^{-1} + \mathbf{I}_{N_{\text{E}}} \right) \right). \quad (23)$$

*Proof:* See Appendix F. ∎

*Proposition 4:* As $N_{\text{A}} \to \infty$, the average secrecy rate for the MIMOME-OFDM system in bits/s/Hz is lower bounded as follows:

$$\mathbb{E}\{R_{\text{sec}}\} \gtrapprox \frac{N_{\text{s}} N}{N + N_{\text{cp}}} \log_2 \left( \frac{\theta \Gamma_{\text{B}}}{N_{\text{s}} N} N_{\text{A}} \tilde{\nu} \sigma_{\text{A-B}}^2 + 1 \right)$$
$$- \frac{N_{\text{E}} N}{N + N_{\text{cp}}} \log_2 \left( \frac{\frac{\theta \Gamma_{\text{E}}}{N} \tilde{\nu} \sigma_{\text{A-E}}^2}{\overline{\theta} \Gamma_{\text{E}} \tilde{\nu} \sigma_{\text{A-E}}^2 \left( \frac{\alpha}{N} + \frac{\overline{\alpha}}{N + N_{\text{cp}}} \right) + 1} + 1 \right) \quad (24)$$

where $\tilde{\nu} = \nu + 1$.

*Proof:* See Appendix G. ∎

The average secrecy rate in (24) is concave in $\alpha$. It is maximized when the denominator of the second term is minimized. The average secrecy rate maximization problem is simplified after removing the terms independent of $\alpha$ as follows:

$$\min_{0 \leq \alpha \leq 1} \frac{\frac{\theta \Gamma_{\mathrm{E}}}{N} \tilde{\nu} \sigma_{\mathrm{A-E}}^2}{\overline{\theta} \Gamma_{\mathrm{E}} \tilde{\nu} \sigma_{\mathrm{A-E}}^2 \left( \frac{\alpha}{N} + \frac{\overline{\alpha}}{N + N_{\mathrm{cp}}} \right) + 1}$$

$$\Rightarrow \max_{0 \leq \alpha \leq 1} \frac{\alpha}{N} + \frac{\overline{\alpha}}{N + N_{\mathrm{cp}}}. \quad (25)$$

Since $N_{\mathrm{cp}}/N \ll 1$, we can approximate $N + N_{\mathrm{cp}}$ by $N$. Hence, the second term in (24) becomes independent of $\alpha$, which implies that $\alpha$ can take any value from 0 to 1 without reducing the average secrecy rate.

*Remark 1:* The lower-bound in (24) holds only when $N_{\mathrm{A}}$ is sufficiently large so that our use of the law of large numbers is accurate. When $N_{\mathrm{A}}$ is not sufficiently large, the bound in (24) should be considered as an approximation.

*Proposition 5:* As $N_{\mathrm{A}} \to \infty$, Eve's average rate when she performs per-subcarrier processing is upper bounded as follows:

$$\mathbb{E}\{R_{\mathrm{A-E},k}\} \lessapprox$$

$$N_{\mathrm{E}} \log_2 \left( \frac{\frac{\theta \Gamma_{\mathrm{E}}}{N} \tilde{\nu} \sigma_{\mathrm{A-E}}^2}{\overline{\theta} \Gamma_{\mathrm{E}} \tilde{\nu} \sigma_{\mathrm{A-E}}^2 \left( \frac{\alpha}{N} + \frac{\overline{\alpha}}{N + N_{\mathrm{cp}}} \right) + 1} + 1 \right). \quad (26)$$

In addition, the average secrecy rate of the legitimate system is lower-bounded as follows:

$$\mathbb{E}\{R_{\mathrm{sec}}\} \gtrapprox \frac{N_{\mathrm{s}} N}{N + N_{\mathrm{cp}}} \log_2 \left( \frac{\theta \Gamma_{\mathrm{B}}}{N_{\mathrm{s}} N} N_{\mathrm{A}} \tilde{\nu} \sigma_{\mathrm{A-B}}^2 + 1 \right)$$

$$- \frac{N_{\mathrm{E}} N}{N + N_{\mathrm{cp}}} \log_2 \left( \frac{\frac{\theta \Gamma_{\mathrm{E}}}{N} \tilde{\nu} \sigma_{\mathrm{A-E}}^2}{\overline{\theta} \Gamma_{\mathrm{E}} \tilde{\nu} \sigma_{\mathrm{A-E}}^2 \left( \frac{\alpha}{N} + \frac{\overline{\alpha}}{N + N_{\mathrm{cp}}} \right) + 1} + 1 \right). \quad (27)$$

*Proof:* Following the same steps as those used to prove Proposition 4, we can prove Proposition 5. ∎

*Proposition 6:* When $N_{\mathrm{A}} \to \infty$ and for high $\Gamma_{\mathrm{B}} \sigma_{\mathrm{A-B}}^2 \geq \Gamma_{\mathrm{E}} \sigma_{\mathrm{A-E}}^2$ and high $\Gamma_{\mathrm{E}} \sigma_{\mathrm{A-E}}^2$, the average secrecy rate for MIMOME-OFDM systems can be lower bounded as follows:

$$\mathbb{E}\{R_{\mathrm{sec}}\} \gtrapprox \frac{N_{\mathrm{s}} N}{N + N_{\mathrm{cp}}} \log_2 \left( \frac{\Gamma_{\mathrm{B}}}{N_{\mathrm{s}} N} N_{\mathrm{A}} \tilde{\nu} \sigma_{\mathrm{A-B}}^2 \right)$$

$$+ \frac{N_{\mathrm{E}} N}{N + N_{\mathrm{cp}}} \log_2 \left( \left( \frac{N_{\mathrm{E}}}{N_{\mathrm{E}} + N_{\mathrm{s}}} \right)^{\frac{N_{\mathrm{s}}}{N_{\mathrm{E}}}} \left( \frac{N_{\mathrm{s}}}{N_{\mathrm{E}} + N_{\mathrm{s}}} \right) \right) \quad (28)$$

when Alice uses the power allocation

$$\theta = \theta^\star = \frac{N_{\mathrm{E}}}{N_{\mathrm{E}} + N_{\mathrm{s}}}. \quad (29)$$

*Proof:* See Appendix H. ∎

Proposition 6 suggests that when the number of transmit antennas at Alice, $N_{\mathrm{A}}$, is large and the SNR is high, the optimal power allocated for data transmission to maximize the average secrecy rate is a function of the number of antennas at Eve and the number of data streams at Alice.

*Corollary 1:* As $N_{\mathrm{A}} \to \infty$ and for high $\Gamma_{\mathrm{B}} \sigma_{\mathrm{A-B}}^2 \geq \Gamma_{\mathrm{E}} \sigma_{\mathrm{A-E}}^2$ and high $\Gamma_{\mathrm{E}} \sigma_{\mathrm{A-E}}^2$, the average secrecy rate loss due to eavesdropping, denoted by $\mathbb{E}\{\mathcal{S}_{\mathrm{Loss}}\}$, in MIMOME-OFDM systems can be upper bounded by

$$\mathbb{E}\{\mathcal{S}_{\mathrm{Loss}}\} \lessapprox \frac{N_{\mathrm{s}} N}{N + N_{\mathrm{cp}}} \log_2 \left( \frac{N_{\mathrm{E}} + N_{\mathrm{s}}}{N_{\mathrm{E}}} \right)$$

$$+ \frac{N_{\mathrm{E}} N}{N + N_{\mathrm{cp}}} \log_2 \left( \frac{N_{\mathrm{E}} + N_{\mathrm{s}}}{N_{\mathrm{s}}} \right) \quad (30)$$

when Alice uses the power allocation

$$\theta = \theta^\star = \frac{N_{\mathrm{E}}}{N_{\mathrm{E}} + N_{\mathrm{s}}}. \quad (31)$$

*Proof:* See Appendix H. ∎

*Remark 2:* From (29), in the high-SNR regime, the optimal power-allocation policy between the data and the AN depends only on the number of receive antennas at Eve and the number of transmitted data streams at Alice. Hence, Alice can optimize $\theta$ that maximizes her average secrecy rate without the need to know the CSI of Eve. If $N_{\mathrm{E}} = N_{\mathrm{s}}$, the optimal power-allocation policy is to set $\theta$ to $1/2$. This implies that the optimal power-allocation policy for Alice is to split her transmit power equally between the data symbols and the AN symbols. If $N_{\mathrm{E}} \gg N_{\mathrm{s}}$, $\theta^\star = 1$, which means that all of Alice's transmit power should be assigned for data transmission to maximize the average secrecy rate, and there is no need to waste power on transmitting AN. Note that if Eve's number of receive antennas is unknown at Alice, Alice can select a value for $\theta$ and operate under this assumption. When Alice does not know any information about Eve, a reasonable value for $\theta$ is $\theta = 1/2$.

*Remark 3:* If $N_{\mathrm{E}} \gg N_{\mathrm{s}}$, the first term of the right-hand side in (30) is approximately zero. Hence

$$\mathbb{E}\{\mathcal{S}_{\mathrm{Loss}}\} \lessapprox \frac{N_{\mathrm{E}} N}{N + N_{\mathrm{cp}}} \log_2 \left( \frac{N_{\mathrm{E}} + N_{\mathrm{s}}}{N_{\mathrm{s}}} \right)$$

$$\approx \frac{N_{\mathrm{E}} N}{N + N_{\mathrm{cp}}} \log_2 \left( \frac{N_{\mathrm{E}}}{N_{\mathrm{s}}} \right). \quad (32)$$

*Proposition 7:* For $N_{\mathrm{E}} = N_{\mathrm{s}}$, $N_{\mathrm{A}} \to \infty$ and for high $\Gamma_{\mathrm{B}} \sigma_{\mathrm{A-B}}^2 \geq \Gamma_{\mathrm{E}} \sigma_{\mathrm{A-E}}^2$ and high $\Gamma_{\mathrm{E}} \sigma_{\mathrm{A-E}}^2$, the average rate loss due to eavesdropping in MIMOME-OFDM systems can be upper bounded as follows:

$$\mathbb{E}\{\mathcal{S}_{\mathrm{Loss}}\} \lessapprox \frac{2 N_{\mathrm{E}} N}{N + N_{\mathrm{cp}}}. \quad (33)$$

*Proof:* Substituting with $N_{\mathrm{s}} = N_{\mathrm{E}}$ into (30), we obtain (33). ∎

*Proposition 8:* As $N_{\mathrm{A}} \to \infty$, $\Gamma_{\mathrm{B}} \sigma_{\mathrm{A-B}}^2 \geq \Gamma_{\mathrm{E}} \sigma_{\mathrm{A-E}}^2$, and for low $\Gamma_{\mathrm{E}} \sigma_{\mathrm{A-E}}^2$, the average secrecy rate loss due to eavesdropping in MIMOME-OFDM systems is upper-bounded as follows:

$$\mathbb{E}\{\mathcal{S}_{\mathrm{Loss}}\} \lessapprox \frac{N_{\mathrm{E}} N}{N + N_{\mathrm{cp}}} \log_2 \left( \frac{\Gamma_{\mathrm{E}}}{N} \tilde{\nu} \sigma_{\mathrm{A-E}}^2 + 1 \right) \quad (34)$$

when Alice uses the power-allocation policy $\theta = \theta^\star = 1$.

*Proof:* See Appendix I. ∎

Proposition 8 suggests that when the number of transmit antennas at Alice is large and the SNR is low, AN injection slightly affects the average secrecy rate. Thus, Alice should allocate all of her power to data transmission (i.e., $\theta = 1$).

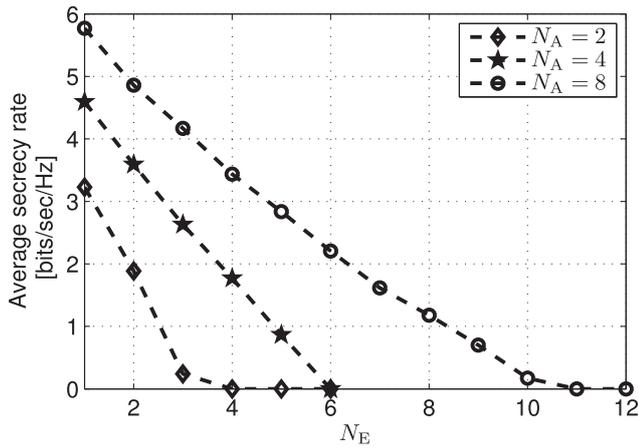

Fig. 2. Average secrecy rate in bits/s/Hz versus $N_E$ for different values of $N_A$.

*Remark 4:* From (89) in Appendix I, when $N_E = N_s = N_B$, the average secrecy rate is lower-bounded as follows:

$$\mathbb{E}\{R_{\text{sec}}\} \gtrapprox \frac{N_s N}{N + N_{\text{cp}}} \log_2\left(\frac{\Gamma_B}{N_s N} N_A \tilde{\nu} \sigma^2_{\text{A-B}} + 1\right) \quad (35)$$
$$- \frac{N_s N}{N + N_{\text{cp}}} \log_2\left(\frac{\Gamma_E}{N} \tilde{\nu} \sigma^2_{\text{A-E}} + 1\right).$$

Since $N_A > N_s$ and $N_A \to \infty$, we have

$$\log_2\left(\frac{\Gamma_B}{N_s N} N_A \tilde{\nu} \sigma^2_{\text{A-B}} + 1\right) \gg \log_2\left(\frac{\Gamma_E}{N} \tilde{\nu} \sigma^2_{\text{A-E}} + 1\right). \quad (36)$$

Consequently, (35) is approximately given by

$$\mathbb{E}\{R_{\text{sec}}\} \gtrapprox \frac{N_s N}{N + N_{\text{cp}}} \log_2\left(\frac{\Gamma_B}{N_s N} N_A \tilde{\nu} \sigma^2_{\text{A-B}} + 1\right) \quad (37)$$
$$= \mathbb{E}\{R_{\text{A-B,noEve}}\}.$$

From (37) and given that $\mathbb{E}\{R_{\text{A-B,noEve}}\} \geq \mathbb{E}\{R_{\text{sec}}\}$, we conclude that at low input power levels and for a large number of transmit antennas at Alice, the average secrecy rate loss is negligible.

### B. Temporal AN Versus Spatial AN

We conclude this section by comparing the pros and cons of spatial and temporal AN for MIMOME-OFDM systems.

1) Since Alice encodes the data on each subcarrier individually, and the received signals at different subcarriers at Bob are independent, Bob can perform per-subcarrier processing of the received signals without loss of optimality. On the other hand, due to the presence of the temporal AN, which couples the signals at different subcarriers (i.e., results in a correlated noise vector at Eve's receiver), Eve's optimal detection strategy should be based on joint processing on the received subcarriers. However, in our numerical results, Eve's rate loss due to per-sub-carrier processing is not significant for the considered scenarios, as shown in Fig. 3. This demonstrates the benefits of injecting the temporal AN to increase the decoding complexity at Eve.

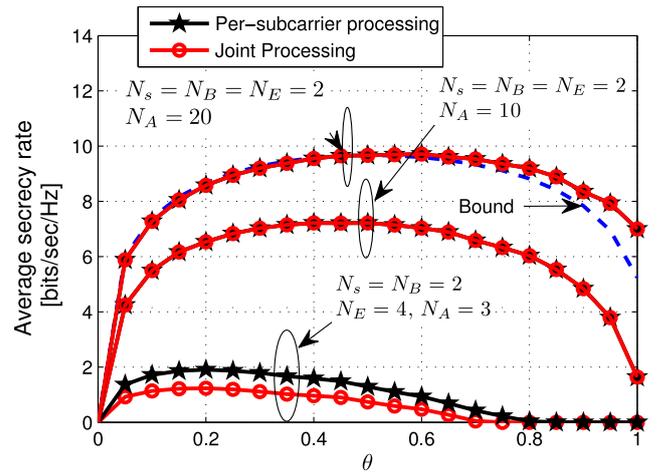

Fig. 3. Average secrecy rate in bits/s/Hz versus $\theta$ for different antenna configurations.

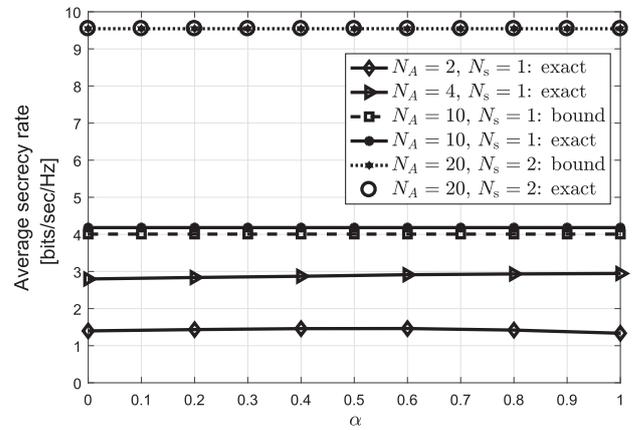

Fig. 4. Average secrecy rate in bits/s/Hz versus $\alpha$.

2) When $N_A = N_s$, the dimension of the AN vector is zero. Hence, spatial AN cannot be applied, and Alice can only apply temporal AN to confuse Eve.
3) The implementation complexity of spatial AN is lower than that of temporal AN. More specifically, the spatial AN precoder is implemented using the SVD of the $N$ per-subcarrier matrices, where the dimension of each subcarrier matrix is $N_s \times N_A$. Hence, the complexity of computing the overall spatial AN precoder matrix **B** is linear in $N$ and cubic in $N_s$ and $N_A$. On the other hand, the temporal AN design requires the computation of the null space of a large matrix whose size is $NN_s \times (N + N_{\text{cp}})N_A$, which is cubic in $N$.
4) Temporal AN injection increases the encoding complexity at Alice as well as the decoding complexity at Eve. On the other hand, spatial AN injection decreases complexity at both nodes. However, since Eve does not know exactly which AN Alice is injecting in her transmissions, Alice might use the spatial AN scheme, while Eve performs joint subcarrier processing.
5) Given that both schemes provide comparable average secrecy rates as proved in Proposition 1 and demonstrated numerically in Fig. 4, Alice can inject the spatial-only AN to reduce the system design complexity. However, spatial AN cannot be implemented when $N_A = N_s$. We emphasize here that our proposed null space computation technique in Appendix A, which is used to compute the temporal AN precoding matrix, reduces the null space computation complexity and can increase the randomness at Eve since Alice can use random matrices in her design. Thus, we conjecture that temporal AN injection can further increase the secrecy rate using a randomized temporal AN precoder matrix.

## IV. SIMULATION RESULTS

In this section, we evaluate the average secrecy rate of our proposed hybrid spatial–temporal AN-aided scheme. We use the following common parameters: $\nu = N_{\rm cp} = 16$, $N = 64$, $\Gamma_B = \Gamma_E = 20$ dB, and $\sigma^2_{\rm A-B} = \sigma^2_{\rm A-E} = 1$. Fig. 2 shows the impact of the number of antennas at Eve on the average secrecy rate. The parameters used to generate the figure are the common parameters and $\theta = \alpha = 0.5$. As $N_E$ increases, Eve can better mitigate the impact of the AN, and hence, the average secrecy rate decreases. When $N_E$ is large enough, Eve can always decode Alice's data reliably, and hence, the secrecy rate is zero. This figure also quantifies the impact of $N_A$ on the average secrecy rate. For example, when $N_E = 4$, the average secrecy rate is increased from 0 to 2 bits/s/Hz as $N_A$ increases from 2 to 4.

Fig. 3 quantifies two important aspects: 1) the impact of the power fraction assigned to information-bearing signal $\theta$ on the average secrecy rate and 2) the accuracy of our proposed approximations and bounds. The parameters used to generate the figure are the common parameters and $\alpha = 0.5$. As the power fraction increases, the average secrecy rate increases till it achieves a maximum; then, it decreases till it achieves the zero secrecy rate since there is no sufficient power to transmit information-bearing signal. If Eve's CSI is known at Alice, she can optimize over $\theta$ for each OFDM block to maximize the instantaneous secrecy rate. If the statistics of the Alice–Eve links are known at Alice, she can optimize $\theta$ to maximize the average secrecy rate. However, if Alice has no information about Eve's channels, she chooses a value for $\theta$ and operates under this assumption.

When $N_A$ is very large, as shown in Section III-A, the optimal $\theta$ depends on the number of transmit data streams at Alice and the number of receive antennas at Eve. Hence, Alice just needs to know the number of receive antennas at Eve to adjust $\theta$ such that the average secrecy rate is maximized. For $N_s = N_B = N_E = 2$ and $N_A = 10$, per-subcarrier processing achieves the same average secrecy rate as that of joint subcarrier processing. This means that Eve can reduce the decoding complexity at her receiver by employing per-subcarrier processing. On the other hand, for the scenario of $N_s = N_B = 2$, $N_E = 4$, and $N_A = 3$, joint processing increases Eve's rate and, hence, decreases the secrecy rate of the legitimate system. We emphasize here that if Alice does not know the CSI of the Alice–Eve links or their statistics, she cannot optimize $\theta$ to maximize her own secrecy rate. However, a good approximation to the optimal $\theta$ is given by (81), e.g., $\theta^\star = N_E/(N_E + N_s) = 1/2$ is optimal for $N_A = 10$ and $N_A = 20$. Eve can still decide to perform per-subcarrier processing even if she achieves a lower rate to reduce the decoding complexity since the secrecy rate loss is not high in many scenarios. Furthermore, since we assume the worst-case scenario, in which Eve knows all channels and precoder matrices, Eve can choose the best decoding strategy for her. Based on our extensive simulations, we conjecture that per-subcarrier processing at Eve achieves a comparable rate as joint processing when $N_A > N_E$. When $N_A = 20$, our proposed lower-bound on the average secrecy rate is tight for low $\theta$ levels. When $\theta$ is near unity, our proposed lower-bound is close to the exact average secrecy rate. Fig. 3 reveals that if Alice allocates all her power to the data (i.e., $\theta = 1$), the average secrecy rate may approach zero as in the scenario of $N_s = N_B = 2$, $N_E = 4$, and $N_A = 3$. This demonstrates the benefits of AN injection.

Fig. 4 demonstrates the impact of the transmit power distribution between the spatial and temporal AN on the average secrecy rate. This figure is generated using the common parameters: $\theta = 0.5$, $N_B = N_s$, and $N_E = 2$. We plot the average secrecy rate by varying $\alpha$, which is the fraction of the AN power allocated to the spatial AN. This figure also shows the secrecy rate increase with the numbers of transmit antennas at Alice and receive antennas at Bob. The impact of $\alpha$ is relatively small and the secrecy rate versus $\alpha$ is almost flat. The benefit of increasing the number of transmit antennas is evident since the secrecy rate is almost doubled when $N_A$ is doubled. The figure also demonstrates the tightness of our proposed lower-bound. The bound completely coincides with the exact value when $N_A = 20$ and $N_s = 2$. For $N_A = 10$ and $N_s = 2$, the bound is very close to the exact value and the gap is less than 5%.

## V. CONCLUSION

In this paper, we proposed a hybrid spatio-temporal AN-aided scheme to secure a MIMOME-OFDM system in the presence of a passive eavesdropper. We assumed that Eve's CSI is unknown at the legitimate nodes. We analyzed the nodes' rates, system's secrecy rate, and system's average secrecy rate and derived lower-bounds on the secrecy rate of the MIMOME-OFDM systems when the number of transmit antennas at Alice, $N_A$, is very large. When $N_A$ is sufficiently large, closed-form solutions for the optimal power allocation between the AN and data and between the temporal and spatial AN were obtained from these bounds at high and low average SNRs.

In the low-SNR regime, Alice should not allocate power to AN, and hence, all Alice's transmit power is allocated to data transmission. The corresponding average secrecy rate loss due to the presence of an eavesdropper is negligible. In the high-SNR regime, Alice should allocate $\theta^\star = N_E/(N_E + N_s)$ of her total power for information transmission and the remaining power for AN transmission. If $N_E = N_s$, in the high average SNR regime, Alice should allocate half of her total power to AN transmission and the other half to data transmission. The corresponding average secrecy rate loss due to the presence of an eavesdropper is $2/(N + N_{\rm cp})$ bits/s/Hz, which is linearly decreasing with the OFDM block size $N$. In addition, we numerically demonstrated that this power allocation solution $\theta^\star = N_E/(N_E + N_s)$ is also a good approximation for the case with a small $N_A$ and can be used without knowledge of the Alice–Eve links CSI at Alice. Furthermore, we showed that the average secrecy rate is independent of $\alpha$ for $N_{cp}/N \ll 1 - N_s/N_A$. In addition, we showed that Eve can perform per-subcarrier decoding instead of joint decoding without significant reduction in her rate.

## APPENDIX A
## TEMPORAL AN PRECODER MATRIX DESIGN AND EFFICIENT COMPUTATION

In this Appendix, we design the temporal AN precoder matrix $\mathbf{Q}$ for the case $N_A \geq N_B = N_s$. From (5), the condition to cancel the temporal AN at Bob when $N_B = N_s$ becomes

$$\mathbf{C}_B^* \mathbf{P}_{N_B}^\top \mathbf{F}_{N_B} \mathbf{R}_{N_B}^{cp} \tilde{\mathbf{H}} \mathbf{Q} = \mathbf{0} \Rightarrow \mathbf{R}_{N_B}^{cp} \tilde{\mathbf{H}} \mathbf{Q} = \mathbf{0} \quad (38)$$

since the matrices $\mathbf{C}_B^*$, $\mathbf{P}_{N_B}^\top$, and $\mathbf{F}_{N_B}$ are square unitary when $N_B = N_s$. The matrix $\tilde{\mathbf{H}}_{toep} = \mathbf{R}_{N_B}^{cp} \tilde{\mathbf{H}}$ in (5) and (38) is a block-Toeplitz matrix whose $(i,j)$ block is the Toeplitz matrix $\tilde{\mathbf{H}}_{i,j}$, which represents the CIR matrix of the $i-j$ link. The matrix $\tilde{\mathbf{H}}_{i,j} \in \mathbb{C}^{(N+N_{cp}) \times (N+N_{cp})}$ can be split into two blocks. The first block, denoted by $\tilde{\mathbf{H}}_{i,j}^1$, is $N \times N$, while the second block, denoted by $\tilde{\mathbf{H}}_{i,j}^2$, has a size of $N \times N_{cp}$. The determinant of $\tilde{\mathbf{H}}_{i,j}^1$ is $a^N$ (where $a$ is the diagonal element of the Toeplitz upper-triangular matrix $\tilde{\mathbf{H}}_{i,j}^1$), and hence, it is a full-rank matrix.

Starting with the special case of $N_A = N_B = 1$, the matrix $\tilde{\mathbf{H}}_{toep} \mathbf{Q}$ is given by

$$\tilde{\mathbf{H}}_{toep} \mathbf{Q} = \begin{pmatrix} \tilde{\mathbf{H}}_{1,1}^1 & \tilde{\mathbf{H}}_{1,1}^2 \end{pmatrix} \underbrace{\begin{pmatrix} \mathbf{Q}_1 \\ \mathbf{Q}_2 \end{pmatrix}}_{=\mathbf{Q}} \quad (39)$$

where the sizes of $\mathbf{Q}_1$ and $\mathbf{Q}_2$ are $N \times N_{cp}$ and $N_{cp} \times N_{cp}$, respectively. The condition to cancel the AN at Bob is

$$\tilde{\mathbf{H}}_{toep} \mathbf{Q} = \mathbf{0} \Rightarrow \mathbf{Q}_1 = -\left(\tilde{\mathbf{H}}_{1,1}^1\right)^{-1} \tilde{\mathbf{H}}_{1,1}^2 \mathbf{Q}_2. \quad (40)$$

Hence, Alice can randomly generate $\mathbf{Q}_2$ and compute $\mathbf{Q}_1$ from (40). Since the matrix $\tilde{\mathbf{H}}_{1,1}^1$ in (40) is an $N \times N$ Toeplitz matrix, when $N$ is large, we can approximate it by a circulant matrix [29], which reduces its inversion complexity significantly by applying the FFT operation. After computing $\mathbf{Q}_1$ and $\mathbf{Q}_2$, we apply the Gram–Schmidt procedure on the columns of $\mathbf{Q} = (\mathbf{Q}_1^\top, \mathbf{Q}_2^\top)^\top$ to make it an orthonormal-column matrix. If $\mathbf{Q}_2$ is generated to have zero-mean i.i.d. complex Gaussian entries with unit variance, $\mathbf{Q}_1$ becomes a matrix with complex zero-mean Gaussian entries.

When Alice has $N_A > 1$ antennas and Bob has only one antenna, i.e., $N_B = N_s = 1$ antenna, the condition to cancel the AN signal at Bob becomes

$$\tilde{\mathbf{H}}_{toep} \mathbf{Q} = \sum_{i=1}^{N_A} \tilde{\mathbf{H}}_{1,i}^1 \mathbf{Q}_{2i+1} + \sum_{i=1}^{N_A} \tilde{\mathbf{H}}_{1,i}^2 \mathbf{Q}_{2i} = \mathbf{0}$$

$$\tilde{\mathbf{H}}_{1,j}^1 \mathbf{Q}_{2j+1} = -\sum_{\substack{i=1 \\ i \neq j}}^{N_A} \tilde{\mathbf{H}}_{1,i}^1 \mathbf{Q}_{2i+1} - \sum_{i=1}^{N_A} \tilde{\mathbf{H}}_{1,i}^2 \mathbf{Q}_{2i}$$

$$\mathbf{Q}_{2j+1} = -\left(\tilde{\mathbf{H}}_{1,j}^1\right)^{-1} \left(\sum_{\substack{i=1 \\ i \neq j}}^{N_A} \tilde{\mathbf{H}}_{1,i}^1 \mathbf{Q}_{2i+1} + \sum_{i=1}^{N_A} \tilde{\mathbf{H}}_{1,i}^2 \mathbf{Q}_{2i}\right)$$
(41)

where $\mathbf{Q} = (\mathbf{Q}_1^\top, \mathbf{Q}_2^\top, \ldots, \mathbf{Q}_{2N_A}^\top)^\top$. From (41), which holds for any $j$, we notice that Alice can randomly generate $\mathbf{Q}_i$, for all $i \neq 2j+1$, and then obtain $\mathbf{Q}_{2j+1}$ from (41) where the inverse of the Toeplitz matrix $\tilde{\mathbf{H}}_{1,j}^1$ is computed approximately using the FFT to reduce computational complexity. Finally, Alice applies the Gram–Schmidt procedure on the columns of $\mathbf{Q}$ to make it an orthonormal-column matrix.

If $N_B > 1$, Alice stacks $N_s$ matrices of size $N \times N$ in a single bigger matrix of size $N_s N \times N_s N$. As an example, consider the case of $N_B = N_s = 2$, where we have

$$\tilde{\mathbf{H}}_{toep} \mathbf{Q} = \begin{pmatrix} \tilde{\mathbf{H}}_{1,1}^1 & \tilde{\mathbf{H}}_{1,1}^2 & \tilde{\mathbf{H}}_{1,2}^1 & \tilde{\mathbf{H}}_{1,2}^2 & \cdots & \tilde{\mathbf{H}}_{1,N_A}^1 & \tilde{\mathbf{H}}_{1,N_A}^2 \\ \tilde{\mathbf{H}}_{2,1}^1 & \tilde{\mathbf{H}}_{2,1}^2 & \tilde{\mathbf{H}}_{2,2}^1 & \tilde{\mathbf{H}}_{2,2}^2 & \cdots & \tilde{\mathbf{H}}_{2,N_A}^1 & \tilde{\mathbf{H}}_{2,N_A}^2 \end{pmatrix}$$

$$\times \begin{pmatrix} \mathbf{Q}_1 \\ \mathbf{Q}_2 \\ \mathbf{Q}_3 \\ \mathbf{Q}_4 \\ \vdots \\ \mathbf{Q}_{2N_A} \end{pmatrix} = \mathbf{0}$$

$$= \begin{pmatrix} \tilde{\mathbf{H}}_{1,1}^1 & \tilde{\mathbf{H}}_{1,2}^1 \\ \tilde{\mathbf{H}}_{2,1}^1 & \tilde{\mathbf{H}}_{2,2}^1 \end{pmatrix} \begin{pmatrix} \mathbf{Q}_1 \\ \mathbf{Q}_3 \end{pmatrix}$$

$$+ \begin{pmatrix} \tilde{\mathbf{H}}_{1,1}^2 & \tilde{\mathbf{H}}_{1,2}^2 & \cdots & \tilde{\mathbf{H}}_{1,N_A}^1 & \tilde{\mathbf{H}}_{1,N_A}^2 \\ \tilde{\mathbf{H}}_{2,1}^2 & \tilde{\mathbf{H}}_{2,2}^2 & \cdots & \tilde{\mathbf{H}}_{2,N_A}^1 & \tilde{\mathbf{H}}_{2,N_A}^2 \end{pmatrix}$$

$$\times \begin{pmatrix} \mathbf{Q}_2 \\ \mathbf{Q}_4 \\ \mathbf{Q}_5 \\ \vdots \\ \mathbf{Q}_{2N_A-1} \\ \mathbf{Q}_{2N_A} \end{pmatrix} = \mathbf{0}.$$
(42)

The matrix $\begin{pmatrix} \tilde{\mathbf{H}}^1_{1,1} & \tilde{\mathbf{H}}^1_{1,2} \\ \tilde{\mathbf{H}}^1_{2,1} & \tilde{\mathbf{H}}^1_{2,2} \end{pmatrix}$ is invertible since its determinant is equal to $\det\left(\tilde{\mathbf{H}}^1_{1,1}\tilde{\mathbf{H}}^1_{2,2} - \tilde{\mathbf{H}}^1_{1,2}\left(\tilde{\mathbf{H}}^1_{2,2}\right)^{-1}\tilde{\mathbf{H}}^1_{2,1}\tilde{\mathbf{H}}^1_{2,2}\right)$ which is nonzero. This is due to the fact that the matrix $\tilde{\mathbf{H}}^1_{1,1}\tilde{\mathbf{H}}^1_{2,2} - \tilde{\mathbf{H}}^1_{1,2}\left(\tilde{\mathbf{H}}^1_{2,2}\right)^{-1}\tilde{\mathbf{H}}^1_{2,1}\tilde{\mathbf{H}}^1_{2,2}$ is an upper-triangular matrix with complex Gaussian random entries, and hence, it is full rank.[5] Rearranging (42), we get

$$\begin{pmatrix} \mathbf{Q}_1 \\ \mathbf{Q}_3 \end{pmatrix} = -\begin{pmatrix} \tilde{\mathbf{H}}^1_{1,1} & \tilde{\mathbf{H}}^1_{1,2} \\ \tilde{\mathbf{H}}^1_{2,1} & \tilde{\mathbf{H}}^1_{2,2} \end{pmatrix}^{-1}$$
$$\times \begin{pmatrix} \tilde{\mathbf{H}}^2_{1,1} & \tilde{\mathbf{H}}^2_{1,2} & \cdots & \tilde{\mathbf{H}}^1_{1,N_A} & \tilde{\mathbf{H}}^2_{1,N_A} \\ \tilde{\mathbf{H}}^2_{2,1} & \tilde{\mathbf{H}}^2_{2,2} & \cdots & \tilde{\mathbf{H}}^1_{2,N_A} & \tilde{\mathbf{H}}^2_{2,N_A} \end{pmatrix} \begin{pmatrix} \mathbf{Q}_2 \\ \mathbf{Q}_4 \\ \mathbf{Q}_5 \\ \vdots \\ \mathbf{Q}_{2N_A-1} \\ \mathbf{Q}_{2N_A} \end{pmatrix}. \quad (43)$$

Alice generates $\mathbf{Q}_i$, for all $i \notin \{1,3\}$, randomly and calculates $\mathbf{Q}_1$ and $\mathbf{Q}_3$ using (43). We can generalize the above approach for any $N_B$ at Bob.

*Remark 5:* Although our proposed algorithm for the null space computation of a Toeplitz upper-triangular matrix has the ability to randomly generate the columns of the temporal AN precoder matrix, which implies that Eve cannot possibly know the precoder, we still assume that Eve can estimate it. This represents the best-case scenario for Eve and gives her an unrealistic ability to know the precoder.

## APPENDIX B
## PROOF OF PROPOSITION 1

In this Appendix, we prove that Eve's rate and the secrecy rate are independent of $0 \leq \alpha \leq 1$ by showing that when $N_{\mathrm{cp}}/N \ll 1 - N_{\mathrm{s}}/N_A$, the eigenvalues of the interference-plus-noise covariance matrix, denoted by $\boldsymbol{\Sigma}_{\mathrm{AN}}$ in (12), do not depend on $\alpha$. Starting from (4) and (5), define $\mathcal{X}_{\tilde{\mathbf{H}}} = \mathbf{C}^*_B \mathbf{P}^\top_{N_B} \mathbf{F}_{N_B} \mathbf{R}^{\mathrm{cp}}_{N_B} \tilde{\mathbf{H}}$. Hence, we can rewrite the terms in (4) and (5) as $\mathbf{C}^*_B \mathbf{HB} = \mathcal{X}_{\tilde{\mathbf{H}}} \mathbf{T}^{\mathrm{cp}}_{N_A} \mathbf{F}^*_{N_A} \mathbf{P}_{N_A} \mathbf{B}$ and $\mathbf{C}^*_B \mathbf{P}^\top_{N_B} \mathbf{F}_{N_B} \mathbf{R}^{\mathrm{cp}}_{N_B} \tilde{\mathbf{H}} \mathbf{Q} = \mathcal{X}_{\tilde{\mathbf{H}}} \mathbf{Q}$. When $N_{\mathrm{cp}}/N \ll 1$, the CP insertion and removal matrices are approximately identity matrices. Hence, the spatial AN cancellation condition in (4) can be rewritten as

$$\mathcal{X}_{\tilde{\mathbf{H}}} \hat{\mathbf{B}} = \mathbf{0} \quad (44)$$

where $\hat{\mathbf{B}} = \mathbf{T}^{\mathrm{cp}}_{N_A} \mathbf{F}^*_{N_A} \mathbf{P}_{N_A} \mathbf{B}$ is an orthonormal-column matrix. The temporal AN cancellation condition in (5) can be rewritten as

$$\mathcal{X}_{\tilde{\mathbf{H}}} \mathbf{Q} = \mathbf{0}. \quad (45)$$

Since the columns of $\mathbf{Q}$ form an orthonormal basis for the null space of $\mathcal{X}_{\tilde{\mathbf{H}}}$ and the columns of $\hat{\mathbf{B}}$ are orthonormal and they lie in the null space of $\mathcal{X}_{\tilde{\mathbf{H}}}$, every column in $\hat{\mathbf{B}}$ can be written as a linear combination of the columns of $\mathbf{Q}$. Hence

$$\mathbf{QL} = \hat{\mathbf{B}} \quad (46)$$

where $\mathbf{L} \in \mathbb{C}^{(N(N_A-N_s)+N_{\mathrm{cp}}N_A) \times N(N_A-N_s)}$ is a linear transformation matrix. Since $\mathbf{Q}^*\mathbf{Q} = \mathbf{I}_{N(N_A-N_s)+N_{\mathrm{cp}}N_A}$ and $\hat{\mathbf{B}}^*\hat{\mathbf{B}} = \mathbf{I}_{N(N_A-N_s)}$, we have

$$\mathbf{L}^*\mathbf{Q}^*\mathbf{QL} = \hat{\mathbf{B}}^*\hat{\mathbf{B}} = \mathbf{I}_{N(N_A-N_s)}. \quad (47)$$

Consequently

$$\mathbf{L}^*\mathbf{L} = \mathbf{I}_{N(N_A-N_s)}. \quad (48)$$

This implies that $\mathbf{L}$ is an orthonormal-column matrix. Substituting (46) into (12), we have

$$\boldsymbol{\Sigma}_{\mathrm{AN}} = \bar{\theta}\Gamma_{\mathrm{E}}\mathcal{X}_{\tilde{\mathbf{G}}}\left(\frac{\alpha \mathbf{QLL}^*\mathbf{Q}^*}{N(N_A-N_s)}\right.$$
$$\left.+ \frac{\bar{\alpha}\mathbf{QQ}^*}{N(N_A-N_s)+N_{\mathrm{cp}}N_A}\right)\mathcal{X}^*_{\tilde{\mathbf{G}}}$$
$$= \bar{\theta}\Gamma_{\mathrm{E}}\mathcal{X}_{\tilde{\mathbf{G}}}\mathbf{Q}\left(\frac{\alpha}{N(N_A-N_s)}\mathbf{LL}^*\right.$$
$$\left.+ \frac{\bar{\alpha}}{N(N_A-N_s)+N_{\mathrm{cp}}N_A}\mathbf{I}_{N(N_A-N_s)+N_{\mathrm{cp}}N_A}\right)\mathbf{Q}^*\mathcal{X}^*_{\tilde{\mathbf{G}}} \quad (49)$$

where $\mathcal{X}_{\tilde{\mathbf{G}}} = \mathbf{P}^\top_{N_E} \mathbf{F}_{N_E} \mathbf{R}^{\mathrm{cp}}_{N_E} \tilde{\mathbf{G}}$. Since $N_{\mathrm{cp}}/N \ll 1$, we have

$$\boldsymbol{\Sigma}_{\mathrm{AN}} = \frac{\bar{\theta}\Gamma_{\mathrm{E}}}{N(N_A-N_s)}\mathbf{D}\left(\alpha\mathbf{LL}^* + \bar{\alpha}\mathbf{I}_{N(N_A-N_s)+N_{\mathrm{cp}}N_A}\right)\mathbf{D}^* \quad (50)$$

where $\mathbf{D} = \mathcal{X}_{\tilde{\mathbf{G}}}\mathbf{Q}$. Define the SVD of $\mathbf{L}$ as $\mathbf{L} = \mathbf{U_L}\boldsymbol{\Lambda_L}\mathbf{V}^*_\mathbf{L}$, where the columns of $\mathbf{U_L}$ are the left singular vectors of $\mathbf{L}$, $\boldsymbol{\Lambda_L} = \mathrm{blkdiag}(\mathbf{I}_{N(N_A-N_s)}, \mathbf{0}) \in \mathbb{C}^{(N(N_A-N_s)+N_{\mathrm{cp}}N_A) \times (N(N_A-N_s)+N_{\mathrm{cp}}N_A)}$ is the diagonal matrix containing the singular values, and $\mathbf{V_L}$ are the right singular vectors of $\mathbf{L}$. Thus

$$\boldsymbol{\Sigma}_{\mathrm{AN}} = \frac{\bar{\theta}\Gamma_{\mathrm{E}}}{N(N_A-N_s)}$$
$$\times \mathbf{D}\left(\alpha\mathbf{U_L}\boldsymbol{\Lambda}^2_\mathbf{L}\mathbf{U}^*_\mathbf{L} + \bar{\alpha}\mathbf{I}_{N(N_A-N_s)+N_{\mathrm{cp}}N_A}\right)\mathbf{D}^*. \quad (51)$$

---

[5] We used the facts that 1) the inverse of an upper triangular matrix is another upper triangular matrix [30] and that 2) the multiplication of two upper triangular matrices is also an upper-triangular matrix [31].

By replacing $\mathbf{I}_{N(N_A-N_s)+N_{cp}N_A}$ in (51) with $\mathbf{U_L U_L^*}$, and since $\mathbf{\Lambda_L^2} = \mathbf{\Lambda_L}$, (51) is rewritten as

$$\begin{aligned}\mathbf{\Sigma}_{AN} &= \frac{\bar{\theta}\Gamma_E}{N(N_A-N_s)}\mathbf{D}\left(\alpha\mathbf{U_L\Lambda_L U_L^*} + \bar{\alpha}\mathbf{U_L U_L^*}\right)\mathbf{D}^* \\ &= \frac{\bar{\theta}\Gamma_E}{N(N_A-N_s)}\mathbf{D U_L} \\ &\quad \times \left(\alpha\mathbf{\Lambda_L} + \bar{\alpha}\mathbf{I}_{N(N_A-N_s)+N_{cp}N_A}\right)\mathbf{U_L^* D^*} \\ &= \bar{\theta}\Gamma_E \mathbf{D U_L} \frac{\mathbf{M}}{N(N_A-N_s)}\mathbf{U_L^* D^*} \end{aligned} \quad (52)$$

where $\mathbf{M} = \alpha\mathbf{\Lambda_L} + \bar{\alpha}\mathbf{I}_{N(N_A-N_s)+N_{cp}N_A} = \mathrm{blkdiag}(\mathbf{I}_{N(N_A-N_s)}, \bar{\alpha}\mathbf{I}_{N_{cp}N_A})$. When the dimension of the matrix $\mathbf{I}_{N(N_A-N_s)}$ is much greater than the dimension of the matrix $\bar{\alpha}\mathbf{I}_{N_{cp}N_A}$, we can ignore the matrix $\bar{\alpha}\mathbf{I}_{N_{cp}N_A}$. The condition is thus given by $N_{cp}/N \ll 1 - N_s/N_A$. In this case, the variation of $\mathbf{\Sigma}_{AN}$ with $\alpha$ is negligible which makes Eve's rate almost flat with $\alpha$. Since the achievable rates at Bob and Eve are independent of $\alpha$, the secrecy rate is also independent of $\alpha$.

## APPENDIX C
## DISTRIBUTION OF $\mathbf{F}_{N_E}\tilde{\mathbf{G}}_{\mathrm{toep}}$

In this Appendix, we derive the distribution of $\mathbf{f}_k\tilde{\mathbf{G}}_{\mathrm{toep}}$ and $\|\mathbf{f}_k\tilde{\mathbf{G}}_{\mathrm{toep}}\|^2$, where $\tilde{\mathbf{G}}_{\mathrm{toep}} = \mathbf{R}_{N_E}^{\mathrm{cp}}\tilde{\mathbf{G}}$ and $\mathbf{f}_k$ is the $k$th row of $\mathbf{F}_{N_E}$. Since the matrix $\tilde{\mathbf{G}}_{\mathrm{toep}}$ is a block Toeplitz matrix whose $(i,j)$ block, denoted by $\tilde{\mathbf{G}}_{i,j}$, is the CIR of the $i-j$ link, and is an upper-triangular Teoplitz matrix with first row equal to $(\tilde{a}, \tilde{b}, \ldots, 0, \ldots, 0)$. This matrix $\tilde{\mathbf{G}}_{ij} \in \mathbb{C}^{(N+N_{cp}) \times (N+N_{cp})}$ has the following properties.
1) Each row has $(\nu+1)$ nonzero circularly symmetric complex Gaussian random variables.
2) The $\ell$th column, $\ell \leq \nu$, has $\ell$ nonzero complex Gaussian entries. The nonzero entries are i.i.d. circuitry-symmetric complex Gaussian.
3) Each column from $(\nu+1)$ to $N$ has $\nu+1$ nonzero complex Gaussian entries.
4) Column $(N+k)$ has $(\nu-k+1)$ nonzero entries. Finally, column $(N+\nu)$ has only 1 nonzero complex Gaussian entry and the other entries are zero.

The above structure is repeated each $N+\nu$ columns and rows of $\tilde{\mathbf{G}}_{\mathrm{toep}}$. The random vector $\mathbf{f}_k\tilde{\mathbf{G}}_{\mathrm{toep}}$ is composed of entries which follow complex Gaussian distribution with different means and variances based on the structure of $\tilde{\mathbf{G}}_{\mathrm{toep}}$. Assuming that $N_s = N_B = N_E = 1$ as in the MISOSE-OFDM case, the $j$th entry of the random vector $\mathbf{f}_k\tilde{\mathbf{G}}_{\mathrm{toep}}$ is given by

$$[\mathbf{f}_k\tilde{\mathbf{G}}_{\mathrm{toep}}]_{1,j} = \sum_{i=1}^{N}[\mathbf{f}_k]_{1,i}\left[\tilde{\mathbf{G}}_{\mathrm{toep}}\right]_{i,j} \quad (53)$$

with $|[\mathbf{f}_k]_{1,i}|^2 = 1/N$ for all $i$. The $j$th entry of $\mathbf{f}_k\tilde{\mathbf{G}}_{\mathrm{toep}}$, $[\mathbf{f}_k\tilde{\mathbf{G}}_{\mathrm{toep}}]_{1,j}$, is a complex Gaussian distributed random variable with zero mean and variance

$$\sigma_j^2 = \begin{cases} \frac{\sigma_{A-E}^2}{N}j, & \text{if } j \leq \nu \\ \frac{\sigma_{A-E}^2}{N}(\nu+1), & \text{if } \nu+1 \leq j \leq N \\ \frac{\sigma_{A-E}^2}{N}(\nu-k+1), & \text{if } j \geq N+k, \forall 1 \leq k \leq \nu \end{cases} \quad (54)$$

where $j \leq N + \nu$. Using (54), the random variable $\|\mathbf{f}_k\tilde{\mathbf{G}}_{\mathrm{toep}}\|^2$ has a mean of

$$\begin{aligned}\mathbb{E}\{\|\mathbf{f}_k\tilde{\mathbf{G}}_{\mathrm{toep}}\|^2\} &= \left(2\frac{\sigma_{A-E}^2}{N} + 2(2)\frac{\sigma_{A-E}^2}{N} + 2(3)\frac{\sigma_{A-E}^2}{N} + \ldots \right.\\ &\quad \left. + 2(\nu)\frac{\sigma_{A-E}^2}{N} + (N-\nu)\tilde{\nu}\frac{\sigma_{A-E}^2}{N}\right)N_A \\ &= \left(2\frac{\sigma_{A-E}^2}{N}\sum_{m=1}^{\nu}m + (N-\nu)\tilde{\nu}\frac{\sigma_{A-E}^2}{N}\right)N_A \\ &= \left(2\frac{\sigma_{A-E}^2}{N}\frac{\nu\tilde{\nu}}{2} + (N-\nu)\tilde{\nu}\frac{\sigma_{A-E}^2}{N}\right)N_A \\ &= \sigma_{A-E}^2\tilde{\nu}N_A \end{aligned} \quad (55)$$

since $\sum_{m=1}^{\nu}m = \frac{\nu(\nu+1)}{2} = \frac{\nu\tilde{\nu}}{2}$. Using the law of large numbers when $N_A \to \infty$, $\|\mathbf{f}_k\tilde{\mathbf{G}}_{\mathrm{toep}}\|^2$ is approximately given by

$$\|\mathbf{f}_k\tilde{\mathbf{G}}_{\mathrm{toep}}\|^2 \approx \sigma_{A-E}^2\tilde{\nu}N_A \quad (56)$$

## APPENDIX D
## DISTRIBUTIONS OF $\mathbf{G}_k\mathbf{A}_k\mathbf{A}_k^*\mathbf{G}_k^*$ AND $\mathbf{H}_k\mathbf{A}_k\mathbf{A}_k^*\mathbf{H}_k^*$

In this Appendix, we derive an analytic expression for the distribution of $\|[\mathbf{G}_k\mathbf{A}_k]_{i,j}\|^2$. The $(i,j)$th element of $\mathbf{G}_k$, $[\mathbf{G}_k]_{i,j}$, is given by

$$[\mathbf{G}_k]_{i,j} = g_{0,i,j} + \sum_{\ell=1}^{\nu}g_{\ell,i,j}\omega^{\ell k} \quad (57)$$

where $\omega = \exp\left(-2\pi\sqrt{-1}/N\right)$, $g_{\ell,i,j}$ is the $\ell$th tap of the CIR of the $(i,j)$th Alice–Eve channel. Since $g_{\ell,i,j}$ follows an i.i.d. Gaussian distribution, $g_{0,i,j} + \sum_{\ell=1}^{\nu}g_{\ell,i,j}\omega^{lk}$ is also Gaussian distributed with zero mean and variance

$$\begin{aligned}&\mathbb{E}\{[\mathbf{G}_k]_{i,j}[\mathbf{G}_k]_{i,j}^*\} \\ &= \mathbb{E}\left\{\left(g_{0,i,j} + \sum_{\ell=1}^{\nu}g_{\ell,i,j}\omega^{\ell k}\right)\left(g_{0,i,j} + \sum_{r=1}^{\nu}g_{r,i,j}\omega^{rk}\right)^*\right\} \\ &= \tilde{\nu}\sigma_{A-E}^2.\end{aligned} \quad (58)$$

Since the channels are i.i.d. random variables with zero mean, the expected values of the cross terms are zero. Each of the diagonal elements of $\mathbf{G}_k\mathbf{G}_k^*$ is the sum of magnitude of squares of $N_A$ Gaussian random variables. Hence, each entry is Chi-square distributed. The off-diagonal entries are the sums of the products of independent complex Gaussian random variables;

hence, their means are zero. As $N_A \to \infty$, the matrix $\mathbf{G}_k \mathbf{G}_k^*$ is approximately given by

$$\mathbf{G}_k \mathbf{G}_k^* \approx \sigma_{A-E}^2 \tilde{\nu} N_A \mathbf{I}_{N_E}. \tag{59}$$

The random variable $[\mathbf{G}_k \mathbf{A}_k \mathbf{A}_k^* \mathbf{G}_k^*]_{i,i}/(\frac{1}{2}\sigma_{A-E}^2 \tilde{\nu})$ is Chi-square distributed with $2N_s$ degrees of freedom since the elements of $\mathbf{G}_k \mathbf{A}_k$ are complex Gaussian random variables with zero mean and variance $\tilde{\nu}\sigma_{A-E}^2$ each. Hence, the expected value of $[\mathbf{G}_k \mathbf{A}_k \mathbf{A}_k^* \mathbf{G}_k^*]_{i,i}$ is $\tilde{\nu}\sigma_{A-E}^2 N_s$. The expected values of the off-diagonal entries of $\mathbf{G}_k \mathbf{A}_k \mathbf{A}_k^* \mathbf{G}_k^*$, $[\mathbf{G}_k \mathbf{A}_k \mathbf{A}_k^* \mathbf{G}_k^*]_{i,j}$, are zeros. In a similar fashion, we can derive the distribution of the random matrix $\mathbf{H}_k \mathbf{A}_k \mathbf{A}_k^* \mathbf{H}_k^*$.

## APPENDIX E
### PROOF OF PROPOSITION 2

When $1 - \frac{N_s}{N_A} \gg \frac{N_{cp}}{N}$, the CP insertion matrix can be approximated by an identity matrix since the CP size is negligible. Moreover, as $N_A \to \infty$, $\mathbf{BB}^* \approx \mathbf{I}_{N(N_A-N_s)}$ and $\mathbf{QQ}^* \approx \mathbf{I}_{N_A(N+N_{cp})}$. Hence, $\mathbf{GG}^* \approx \mathbf{F}_{N_E} \mathbf{R}_{N_E}^{cp} \tilde{\mathbf{G}} (\mathbf{F}_{N_E} \mathbf{R}_{N_E}^{cp} \tilde{\mathbf{G}})^*$ and the covariance matrix of the interference at Eve in (20) becomes

$$\mathbf{\Sigma}_{AN} = \bar{\theta}\Gamma_E \left( \frac{\alpha \mathbf{GG}^*}{N(N_A - N_s)} + \frac{\bar{\alpha} \mathbf{GG}^*}{N(N_A - N_s) + N_{cp}N_A} \right)$$

$$= \bar{\theta}\Gamma_E \mathbf{GG}^* \left( \frac{\alpha}{N(N_A - N_s)} + \frac{\bar{\alpha}}{N(N_A - N_s) + N_{cp}N_A} \right)$$

$$\approx \frac{\bar{\theta}\Gamma_E}{N(N_A - N_s)} \mathbf{GG}^* \tag{60}$$

where the last equality holds when $1 - \frac{N_s}{N_A} \gg \frac{N_{cp}}{N}$. Substituting with $\mathbf{\Sigma}_{AN}$ in (60) into (19) leads to (21). Moreover, substituting with $R_{A-B}$ in (10) and $R_{A-E}$ in (21) into (15), we obtain the secrecy rate in (22), which is independent of $\alpha$.

## APPENDIX F
### PROOF OF PROPOSITION 3

Since $\mathbf{GA}\mathbf{\Sigma}_x (\mathbf{GA})^*$ and $\mathbf{GG}^*$ in (21) are block-diagonal matrices, Eve's rate in (21) can be rewritten as

$$R_{A-E} = \sum_{k=1}^{N} \left[ \log_2 \det \left( \theta \Gamma_E \mathbf{G}_k \mathbf{A}_k \mathbf{\Sigma}_{x_k} (\mathbf{G}_k \mathbf{A}_k)^* \right. \right.$$
$$\left. \left. \times \left( \frac{\bar{\theta}\Gamma_E}{N(N_A - N_s)} \mathbf{G}_k \mathbf{G}_k^* + \mathbf{I}_{N_E} \right)^{-1} + \mathbf{I}_{N_E} \right) \right] \tag{61}$$

which is equivalent to processing each subcarrier individually. This completes the proof.

## APPENDIX G
### PROOF OF PROPOSITION 4

Using Appendixes C and D, we approximate the matrices in (19) and (20) when $N_A$ is very large as follows.

1) From Appendix D, we showed that $\|[\mathbf{G}_k \mathbf{A}_k]_{i,i}\|^2 / (\frac{1}{2}\sigma_{A-E}^2 \tilde{\nu})$ follows the Chi-square distribution with $2N_s$ degrees of freedom, and the mean of $\|[\mathbf{G}_k \mathbf{A}_k]_{i,i}\|^2$ is $\mathbb{E}\{\|[\mathbf{G}_k \mathbf{A}_k]^2\|_{i,i}\} = \tilde{\nu}\sigma_{A-E}^2 N_s$.

2) The matrix $\mathbf{GA} = \text{blkdiag}(\mathbf{G}_1 \mathbf{A}_1, \ldots, \mathbf{G}_N \mathbf{A}_N)$ has diagonal entries whose variances are $N_s \tilde{\nu} \sigma_{A-E}^2$. The matrix $\mathbf{GA}(\mathbf{GA})^*$ has off-diagonal elements $[\mathbf{GA}(\mathbf{GA})^*]_{i,j} = \mathbf{g}_i \mathbf{A}_k \mathbf{A}_k^* \mathbf{g}_j^*$, which represent the sum of $N_s$ i.i.d. complex Gaussian products, where $\mathbf{g}_i$ is the $i$th row of the matrix $\mathbf{G}_k$. The $i$th diagonal element of $\mathbf{GA}(\mathbf{GA})^*$ is $[\mathbf{GA}(\mathbf{GA})^*]_{i,i} = \mathbf{g}_i \mathbf{A}_k \mathbf{A}_k^* \mathbf{g}_i^*$, which follows a Chi-square distribution with mean $\sigma_{A-E}^2 \tilde{\nu} N_s$.

3) *Consider the matrix* $\mathbf{GG}^*$: The matrix $\mathbf{G}$ is a block-diagonal matrix and, hence, $\mathbf{GG}^*$ is also a block-diagonal matrix. The expected value of each of the diagonal entries of $\mathbf{GG}^*$ is equal to $N_A \tilde{\nu} \sigma_{A-E}^2$. The off-diagonal entries are the sum of the product of independent complex Gaussian random variables; hence, their means are zero. Hence, $\mathbf{GG}^*$ is approximated as $N_A \tilde{\nu} \sigma_{A-E}^2 \mathbf{I}_{N_E N}$.

4) Each row of $\mathbf{R}_{N_E}^{cp} \tilde{\mathbf{G}}$ is composed of $\nu + 1$ i.i.d. complex Gaussian random variables. Hence, the expected values of the off-diagonal elements of the matrix $\mathbf{R}_{N_E}^{cp} \tilde{\mathbf{G}} (\mathbf{R}_{N_E}^{cp} \tilde{\mathbf{G}})^* = \mathbf{R}_{N_E}^{cp} \tilde{\mathbf{G}} \tilde{\mathbf{G}}^* \mathbf{R}_{N_E}^{cp*}$ are zero. As $N_A \to \infty$, we can approximate this matrix by $N_A \tilde{\nu} \sigma_{A-E}^2 \mathbf{I}_{N_E N}$. Hence, $\mathbf{P}_{N_E}^\top \mathbf{F}_{N_E} \mathbf{R}_{N_E}^{cp} \tilde{\mathbf{G}} \tilde{\mathbf{G}}^* \mathbf{R}_{N_E}^{cp*} \mathbf{F}_{N_E}^* \mathbf{P}_{N_E} \approx N_A \tilde{\nu} \sigma_{A-E}^2 \mathbf{I}_{N_E N}$ as $N_A \to \infty$.

Based on our earlier discussions, we have the following relations:

$$\mathbb{E}\left\{ [\mathbf{GA}\mathbf{\Sigma}_x (\mathbf{GA})^*]_{i,i} \right\} = \frac{\tilde{\nu}\sigma_{A-E}^2}{N} \tag{62}$$

where $\mathbf{\Sigma}_x = \frac{1}{N_s N} \mathbf{I}_{N_s N}$, and

$$\mathbf{\Sigma}_{AN} = \bar{\theta}\Gamma_E N_A \tilde{\nu} \sigma_{A-E}^2 \lambda_\alpha \mathbf{I}_{N_E N} \tag{63}$$

where $\lambda_\alpha = \left( \frac{\alpha}{N(N_A-N_s)} + \frac{\bar{\alpha}}{N(N_A-N_s) + N_{cp}N_A} \right)$. Hence

$$\mathbf{\Sigma}_{AN} + \mathbf{I}_{N_E N} = \left[ \bar{\theta}\Gamma_E N_A \tilde{\nu} \sigma_{A-E}^2 \lambda_\alpha + 1 \right] \mathbf{I}_{N_E N}. \tag{64}$$

Define a scalar $p(\alpha)$ as

$$p(\alpha) = \bar{\theta}\Gamma_E N_A \tilde{\nu} \sigma_{A-E}^2 \lambda_\alpha + 1. \tag{65}$$

The inverse of $\mathbf{\Sigma}_{AN} + \mathbf{I}_{N_E N}$ is $\frac{1}{p(\alpha)} \mathbf{I}_{N_E N}$. Hence

$$\theta \Gamma_E \mathbf{GA}\mathbf{\Sigma}_x (\mathbf{GA})^* (\mathbf{\Sigma}_{AN} + \mathbf{I}_{N_E N})^{-1} + \mathbf{I}_{N_E N}$$
$$= \frac{\theta \Gamma_E}{p(\alpha)} \mathbf{GA}\mathbf{\Sigma}_x (\mathbf{GA})^* + \mathbf{I}_{N_E N}. \tag{66}$$

Using the Hadamard inequality [32], we can upper bound Eve's rate in (19) as follows:

$$\log_2 \det \left( \theta\Gamma_E \mathbf{GA}\mathbf{\Sigma}_x (\mathbf{GA})^* (\mathbf{\Sigma}_{AN} + \mathbf{I}_{N_E N})^{-1} + \mathbf{I}_{N_E N} \right)$$
$$\leq \log_2 \prod_{i=1}^{N_E N} \left( \frac{\theta\Gamma_E}{p(\alpha)} [\mathbf{GA}\mathbf{\Sigma}_x (\mathbf{GA})^*]_{i,i} + 1 \right). \tag{67}$$

Using the properties of the logarithmic function

$$R_{\mathrm{A-E}} \leq \sum_{i=1}^{N_{\mathrm{E}}N} \log_2\left(\frac{\theta\Gamma_{\mathrm{E}}}{p(\alpha)}[\mathbf{GA\Sigma_x(GA)^*}]_{i,i} + 1\right). \quad (68)$$

The upper bound in (68) is the sum of $N_{\mathrm{E}}N$ concave functions; hence, by using Jensen's inequality, Eve's average rate is upper bounded by

$$\mathbb{E}\{R_{\mathrm{A-E}}\} \leq \sum_{i=1}^{N_{\mathrm{E}}N} \log_2\left(\frac{\theta\Gamma_{\mathrm{E}}}{p(\alpha)}\mathbb{E}\left\{[\mathbf{GA\Sigma_x(GA)^*}]_{i,i}\right\} + 1\right). \quad (69)$$

Using (62), we obtain

$$\mathbb{E}\{R_{\mathrm{A-E}}\} \lessapprox \sum_{i=1}^{N_{\mathrm{E}}N} \log_2\left(\frac{\theta\Gamma_{\mathrm{E}}}{p(\alpha)}\frac{\tilde{\nu}\sigma_{\mathrm{A-E}}^2}{N} + 1\right)$$
$$= N_{\mathrm{E}}N\log_2\left(\frac{\theta\Gamma_{\mathrm{E}}}{p(\alpha)}\frac{\tilde{\nu}\sigma_{\mathrm{A-E}}^2}{N} + 1\right). \quad (70)$$

Substituting for $p(\alpha)$ from (65) into (70) and using the fact that $N_{\mathrm{A}} \gg N_{\mathrm{s}}$, the upper bound on Eve's average rate is given by

$$\mathbb{E}\{R_{\mathrm{A-E}}\} \lessapprox N_{\mathrm{E}}N\log_2\left(\frac{\frac{\theta\Gamma_{\mathrm{E}}}{N}\tilde{\nu}\sigma_{\mathrm{A-E}}^2}{\overline{\theta}\Gamma_{\mathrm{E}}\tilde{\nu}\sigma_{\mathrm{A-E}}^2\left(\frac{\alpha}{N} + \frac{\overline{\alpha}}{N+N_{\mathrm{cp}}}\right) + 1} + 1\right). \quad (71)$$

Since $N$ is much larger than $N_{\mathrm{cp}}$, $N + N_{\mathrm{cp}} \approx N$.[6] Hence, the right-hand side of (71) becomes independent of $\alpha$. This implies that the average Eve's rate is not a function of $\alpha$. Hence

$$\mathbb{E}\{R_{\mathrm{A-E}}\} \lessapprox N_{\mathrm{E}}N\log_2\left(\frac{\frac{\theta\Gamma_{\mathrm{E}}}{N}\tilde{\nu}\sigma_{\mathrm{A-E}}^2}{\frac{\overline{\theta}\Gamma_{\mathrm{E}}\tilde{\nu}\sigma_{\mathrm{A-E}}^2}{N} + 1} + 1\right). \quad (72)$$

The rate of the Alice–Bob link is given by

$$R_{\mathrm{A-B}} = \log_2\det\left(\theta\Gamma_{\mathrm{B}}\mathbf{HA\Sigma_x(HA)^*} + \mathbf{I}_{N_{\mathrm{B}}N}\right). \quad (73)$$

The matrix $\mathbf{HA\Sigma_x(HA)^*}$ can be approximated by its diagonal elements as a weighted identity matrix. Hence, $\mathbf{HA\Sigma_x(HA)^*} \approx \frac{1}{N_{\mathrm{s}}N}N_{\mathrm{A}}\tilde{\nu}\sigma_{\mathrm{A-B}}^2\mathbf{I}_{N_{\mathrm{s}}N}$. Consequently, Bob's average rate can be approximated by

$$\mathbb{E}\{R_{\mathrm{A-B}}\} \approx \log_2\det\left(\frac{\theta\Gamma_{\mathrm{B}}}{N_{\mathrm{s}}N}N_{\mathrm{A}}\tilde{\nu}\sigma_{\mathrm{A-B}}^2\mathbf{I}_{N_{\mathrm{s}}N} + \mathbf{I}_{N_{\mathrm{s}}N}\right)$$
$$= \log_2\left(\frac{\theta\Gamma_{\mathrm{B}}}{N_{\mathrm{s}}N}N_{\mathrm{A}}\tilde{\nu}\sigma_{\mathrm{A-B}}^2 + 1\right)^{N_{\mathrm{s}}N}$$
$$= N_{\mathrm{s}}N\log_2\left(\frac{\theta\Gamma_{\mathrm{B}}}{N_{\mathrm{s}}N}N_{\mathrm{A}}\tilde{\nu}\sigma_{\mathrm{A-B}}^2 + 1\right). \quad (74)$$

Using (71) and (74), and since $\mathbb{E}\{R_{\mathrm{sec}}\} = \mathbb{E}\{R_{\mathrm{A-B}}\} - \mathbb{E}\{R_{\mathrm{A-E}}\}$, we obtain the result in (24).

---

[6]If we did not use the approximation that $N + N_{\mathrm{cp}} \approx N$ when we allocated the powers to data and spatial-AN symbols as discussed in Section II-B, the term $\alpha/N$ in (71) will be replaced with $\alpha/(N + N_{\mathrm{cp}})$. Hence, we will not need the assumption that $N \gg N_{\mathrm{cp}}$ at this point of the proof to show the independence of the upper-bound on Eve's rate on $\alpha$. This remark is also valid for the expression in Eqn. (75).

## APPENDIX H
## PROOF OF PROPOSITION 6

When $\Gamma_{\mathrm{B}}\sigma_{\mathrm{A-B}}^2$ and $\Gamma_{\mathrm{E}}\sigma_{\mathrm{A-E}}^2$ are sufficiently high, we have

$$\mathbb{E}\{R_{\mathrm{sec}}\} \gtrapprox N_{\mathrm{s}}N\log_2\left(\frac{\theta\Gamma_{\mathrm{B}}}{N_{\mathrm{s}}N}N_{\mathrm{A}}\tilde{\nu}\sigma_{\mathrm{A-B}}^2\right)$$
$$- N_{\mathrm{E}}N\log_2\left(\frac{\frac{\theta}{N}}{\overline{\theta}\left(\frac{\alpha}{N} + \frac{\overline{\alpha}}{N+N_{\mathrm{cp}}}\right)} + 1\right). \quad (75)$$

Letting $K_\alpha = N\left(\frac{\alpha}{N} + \frac{\overline{\alpha}}{N+N_{\mathrm{cp}}}\right)$,[7] we lower-bound the average secrecy rate as follows:

$$\mathbb{E}\{R_{\mathrm{sec}}\} \gtrapprox N_{\mathrm{s}}N\log_2\left(\frac{\theta\Gamma_{\mathrm{B}}}{N_{\mathrm{s}}N}N_{\mathrm{A}}\tilde{\nu}\sigma_{\mathrm{A-B}}^2\right)$$
$$- N_{\mathrm{E}}N\log_2\left(\frac{\theta}{\overline{\theta}K_\alpha} + 1\right). \quad (76)$$

Thus

$$\mathbb{E}\{R_{\mathrm{sec}}\} \gtrapprox N_{\mathrm{s}}N\log_2\left(\frac{\Gamma_{\mathrm{B}}}{N_{\mathrm{s}}N}N_{\mathrm{A}}\tilde{\nu}\sigma_{\mathrm{A-B}}^2\right) + N_{\mathrm{E}}N\log_2\left(\theta^{\frac{N_{\mathrm{s}}}{N_{\mathrm{E}}}}\right)$$
$$+ N_{\mathrm{E}}N\log_2\left(\frac{\overline{\theta}K_\alpha}{\theta + \overline{\theta}K_\alpha}\right)$$
$$= N_{\mathrm{s}}N\log_2\left(\frac{\Gamma_{\mathrm{B}}}{N_{\mathrm{s}}N}N_{\mathrm{A}}\tilde{\nu}\sigma_{\mathrm{A-B}}^2\right)$$
$$+ N_{\mathrm{E}}N\log_2\left(\theta^{\frac{N_{\mathrm{s}}}{N_{\mathrm{E}}}}\frac{\overline{\theta}K_\alpha}{\theta + \overline{\theta}K_\alpha}\right). \quad (77)$$

The first term is independent of $\theta$; hence, to increase the lower-bound, Alice needs to choose the value of $\theta$, which maximizes the second term only. This is equivalent to maximize the term inside the logarithmic function. Let

$$f(\theta) = \theta^{\frac{N_{\mathrm{s}}}{N_{\mathrm{E}}}}\frac{\overline{\theta}K_\alpha}{\theta + \overline{\theta}K_\alpha}. \quad (78)$$

When $N \gg N_{\mathrm{cp}}$, $K_\alpha \approx 1$. Hence

$$f(\theta) \approx \theta^{\frac{N_{\mathrm{s}}}{N_{\mathrm{E}}}}\overline{\theta} = \theta^{\frac{N_{\mathrm{s}}}{N_{\mathrm{E}}}} - \theta^{\frac{N_{\mathrm{s}}}{N_{\mathrm{E}}}+1}. \quad (79)$$

The first derivative of $f(\theta)$ is given by

$$\frac{\delta f(\theta)}{\delta\theta} = \frac{N_{\mathrm{s}}}{N_{\mathrm{E}}}\theta^{\frac{N_{\mathrm{s}}}{N_{\mathrm{E}}}-1}\overline{\theta} - \theta^{\frac{N_{\mathrm{s}}}{N_{\mathrm{E}}}}. \quad (80)$$

The root of the first derivative is given by

$$\theta^\star = \frac{1}{1 + \frac{N_{\mathrm{s}}}{N_{\mathrm{E}}}} = \frac{N_{\mathrm{E}}}{N_{\mathrm{E}} + N_{\mathrm{s}}}. \quad (81)$$

The second derivative is given by

$$\frac{\delta^2 f(\theta)}{\delta\theta^2} = \frac{N_{\mathrm{s}}}{N_{\mathrm{E}}}\left(\frac{N_{\mathrm{s}}}{N_{\mathrm{E}}} - 1\right)\theta^{\frac{N_{\mathrm{s}}}{N_{\mathrm{E}}}-2}$$
$$- \left(\frac{N_{\mathrm{s}}}{N_{\mathrm{E}}} + 1\right)\frac{N_{\mathrm{s}}}{N_{\mathrm{E}}}\theta^{\frac{N_{\mathrm{s}}}{N_{\mathrm{E}}}-1} \leq 0 \quad (82)$$

---

[7]If we did not use the approximation that $N + N_{\mathrm{cp}} \approx N$ when we allocated the powers to data and spatial-AN symbols, $K_\alpha = 1$.

since $N_\text{E} \geq N_\text{s}$, which is a reasonable assumption to enable Eve to decode the data.

Substituting with the optimal value of $\theta$ in (81) into (77), the average secrecy rate in bits/s/Hz is lower bounded as follows:

$$\mathbb{E}\{R_\text{sec}\} \gtrapprox N_\text{s} N \log_2\left(\frac{\Gamma_\text{B}}{N_\text{s} N} N_\text{A} \tilde{\nu} \sigma_{\text{A}-\text{B}}^2\right)$$
$$+ N_\text{E} N \log_2\left(\left(\frac{N_\text{E}}{N_\text{E}+N_\text{s}}\right)^{\frac{N_\text{s}}{N_\text{E}}}\left(1 - \frac{N_\text{E}}{N_\text{E}+N_\text{s}}\right)\right). \quad (83)$$

Rearranging the expression in (83), we obtain the result in (28).

When there is no eavesdropper in the network, the average secrecy rate (i.e., average rate of the Alice-Bob link) is

$$\mathbb{E}\{R_{\text{A}-\text{B,noEve}}\} = N_\text{s} N \log_2\left(\frac{\Gamma_\text{B}}{N_\text{s} N} N_\text{A} \tilde{\nu} \sigma_{\text{A}-\text{B}}^2 + 1\right)$$
$$\approx N_\text{s} N \log_2\left(\frac{\Gamma_\text{B}}{N_\text{s} N} N_\text{A} \tilde{\nu} \sigma_{\text{A}-\text{B}}^2\right). \quad (84)$$

The average secrecy rate loss due to the presence of Eve, denoted by $\mathbb{E}\{\mathcal{S}_\text{Loss}\}$, is thus given by

$$\mathbb{E}\{\mathcal{S}_\text{Loss}\} = \mathbb{E}\{R_{\text{A}-\text{B,noEve}}\} - \mathbb{E}\{R_\text{sec}\}$$
$$\lessapprox N_\text{s} N \log_2\left(\frac{\Gamma_\text{B}}{N_\text{s} N} N_\text{A} \tilde{\nu} \sigma_{\text{A}-\text{B}}^2\right)$$
$$- N_\text{s} N \log_2\left(\frac{\Gamma_\text{B}}{N_\text{s} N} N_\text{A} \tilde{\nu} \sigma_{\text{A}-\text{B}}^2\right)$$
$$- N_\text{E} N \log_2\left(\left(\frac{N_\text{E}}{N_\text{E}+N_\text{s}}\right)^{\frac{N_\text{s}}{N_\text{E}}}\left(1 - \frac{N_\text{E}}{N_\text{E}+N_\text{s}}\right)\right). \quad (85)$$

Rearranging (85), we obtain

$$\mathbb{E}\{\mathcal{S}_\text{Loss}\} \lessapprox N_\text{s} N \log_2\left(\frac{N_\text{E}+N_\text{s}}{N_\text{E}}\right) + N_\text{E} N \log_2\left(\frac{N_\text{E}+N_\text{s}}{N_\text{s}}\right). \quad (86)$$

Hence, the average secrecy rate loss in bits/s/Hz is given by the expression in (30).

## APPENDIX I
## PROOF OF PROPOSITION 8

In the low $\Gamma_\text{E} \sigma_{\text{A}-\text{E}}^2 \ll 1$ regime, $\overline{\theta}\Gamma_\text{E}\left(\frac{\alpha}{N}\tilde{\nu}\sigma_{\text{A}-\text{E}}^2 + \frac{\overline{\alpha}}{N+N_\text{cp}}\tilde{\nu}\sigma_{\text{A}-\text{E}}^2\right) \approx \overline{\theta}\frac{\Gamma_\text{E}}{N}\tilde{\nu}\sigma_{\text{A}-\text{E}}^2 \ll 1$. Hence, the upper-bound on Eve's average rate in (72) is approximated as follows:

$$\mathbb{E}\{R_{\text{A}-\text{E}}\} \lessapprox N_\text{E} N \log_2\left(\frac{\frac{\theta\Gamma_\text{E}}{N}\tilde{\nu}\sigma_{\text{A}-\text{E}}^2}{\frac{\overline{\theta}\Gamma_\text{E}}{N}\tilde{\nu}\sigma_{\text{A}-\text{E}}^2 + 1} + 1\right)$$
$$= N_\text{E} N \log_2\left(\frac{\frac{\Gamma_\text{E}}{N}\tilde{\nu}\sigma_{\text{A}-\text{E}}^2 + 1}{\overline{\theta}\frac{\Gamma_\text{E}}{N}\tilde{\nu}\sigma_{\text{A}-\text{E}}^2 + 1}\right) \quad (87)$$
$$\approx N_\text{E} N \log_2\left(\frac{\Gamma_\text{E}}{N}\tilde{\nu}\sigma_{\text{A}-\text{E}}^2 + 1\right).$$

Consequently

$$\mathbb{E}\{R_\text{sec}\} \gtrapprox N_\text{s} N \log_2\left(\frac{\theta\Gamma_\text{B}}{N_\text{s} N} N_\text{A} \tilde{\nu} \sigma_{\text{A}-\text{B}}^2 + 1\right)$$
$$- N_\text{E} N \log_2\left(\frac{\Gamma_\text{E}}{N}\tilde{\nu}\sigma_{\text{A}-\text{E}}^2 + 1\right). \quad (88)$$

From (88), the lower-bound on the average secrecy rate is maximized when $\theta = 1$. Hence

$$\mathbb{E}\{R_\text{sec}\} \gtrapprox N_\text{s} N \log_2\left(\frac{\Gamma_\text{B}}{N_\text{s} N} N_\text{A} \tilde{\nu} \sigma_{\text{A}-\text{B}}^2 + 1\right)$$
$$- N_\text{E} N \log_2\left(\frac{\Gamma_\text{E}}{N}\tilde{\nu}\sigma_{\text{A}-\text{E}}^2 + 1\right). \quad (89)$$

When there is no eavesdropper in the network, the average secrecy rate is

$$\mathbb{E}\{R_{\text{A}-\text{B,noEve}}\} = N_\text{s} N \log_2\left(\frac{\Gamma_\text{B}}{N_\text{s} N} N_\text{A} \tilde{\nu} \sigma_{\text{A}-\text{B}}^2 + 1\right). \quad (90)$$

Accordingly, in the low $\Gamma_\text{E}\sigma_{\text{A}-\text{E}}^2 \leq \Gamma_\text{B}\sigma_{\text{A}-\text{B}}^2$ regime, the average secrecy rate loss is given by

$$\mathbb{E}\{\mathcal{S}_\text{Loss}\} \lessapprox N_\text{E} N \log_2\left(\frac{\Gamma_\text{E}}{N}\tilde{\nu}\sigma_{\text{A}-\text{E}}^2 + 1\right). \quad (91)$$

The average secrecy loss in bits/s/Hz is given by

$$\mathbb{E}\{\mathcal{S}_\text{Loss}\} \lessapprox \frac{N_\text{E} N}{N+N_\text{cp}} \log_2\left(\frac{\Gamma_\text{E}}{N}\tilde{\nu}\sigma_{\text{A}-\text{E}}^2 + 1\right). \quad (92)$$

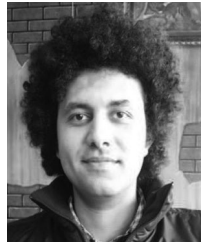

**Ahmed El Shafie** received the B.Sc. degree (with accumulative grade of distinction with honor) in electrical engineering from Alexandria University, Alexandria, Egypt, in 2009 and the M.Sc. degree in communication and information technology from Nile University, Giza, Egypt, in 2014. He is currently working toward the Ph.D. degree with the University of Texas at Dallas, Dallas, TX, USA.

El Shafie was the Exemplary Reviewer for the IEEE TRANSACTIONS ON COMMUNICATIONS in 2015.

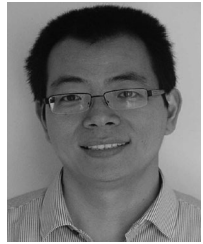

**Zhiguo Ding** (S'03–M'05–SM'15) received the B.Eng. degree in electrical engineering from Beijing University of Posts and Telecommunications, Beijing, China, in 2000 and the Ph.D. degree in electrical engineering from Imperial College London, London, U.K., in 2005.

From July 2005 to August 2014, he was with Queen's University Belfast, Belfast, U.K.; Imperial College; and Newcastle University, Newcastle, U.K.. Since September 2014, he has been with Lancaster University, Lancaster, U.K., as a Chair Professor. From October 2012 to September 2016, he was also an Academic Visitor with Princeton University, Princeton, NJ, USA. His research interests are fifth-generation networks, game theory, cooperative and energy harvesting networks, and statistical signal processing.

Dr. Ding serves as an Editor for the IEEE TRANSACTIONS ON COMMUNICATIONS, the IEEE TRANSACTIONS ON VEHICULAR TECHNOLOGY, IEEE WIRELESS COMMUNICATION LETTERS, IEEE COMMUNICATION LETTERS, and the *Journal of Wireless Communications and Mobile Computing*. He received the Best Paper Award at the IET Communications Conference on Wireless, Mobile, and Computing in 2009, the IEEE Communication Letter Exemplary Reviewer award in 2012, and the EU Marie Curie Fellowship for 2012–2014.

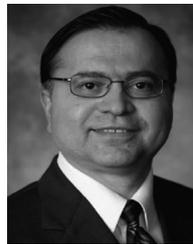

**Naofal Al-Dhahir** (F'07) received the Ph.D. degree in electrical engineering from Stanford University, Stanford, CA, USA, in 1994.

He is the Erik Jonsson Distinguished Professor with the University of Texas at Dallas, Dallas, TX, USA. From 1994 to 2003, he was a Principal Member of the technical staff with GE Research and AT&T Shannon Laboratory. He is co-inventor of 41 issued U.S. patents and coauthor of more than 325 papers with more than 7800 citations.

He has co-received four IEEE Best Paper Awards, including the 2006 IEEE Donald G. Fink Award. He is the Editor-in-Chief of the IEEE TRANSACTIONS ON COMMUNICATIONS.